\documentclass [11pt, revtex-2]{article}

\usepackage{amsmath,amssymb,epsfig,cite,color,verbatim}
\usepackage{graphics,float}
\usepackage{overpic} 
\usepackage{mathrsfs}
\usepackage{soul}
\usepackage{slashed,bbold}
\usepackage{soul}

\numberwithin{equation}{section}

\usepackage[format=plain,
  justification=RaggedRight, 
  singlelinecheck=false, 
  figurename=Fig., 
  aboveskip=15pt, 
  belowskip=0pt]{caption}
  
\title{
Localization-delocalization transition at weak coupling in two-color matrix QCD
}
\date{\today}
\author{Nirmalendu Acharyya$^1$\footnote{nirmalendu@iitbbs.ac.in}, Prasanjit Aich$^{2}$\footnote{prasanjita@iisc.ac.in}, Arkajyoti Bandyopadhyay$^1$\footnote{s22ph09003@iitbbs.ac.in} \\and Sachindeo~Vaidya$^2$\footnote{vaidya@iisc.ac.in} \\
${}^1${\small School of Basic Sciences, Indian Institute of Technology Bhubaneswar, Jatni, Khurda, Odisha 752050, India}\\
${}^2${\small Centre for High Energy Physics,  Indian Institute of Science, Bengaluru, 560012, India}\\
}

\begin{document}
\maketitle
\begin{abstract}
We numerically investigate the matrix model of two-color one-flavor adjoint QCD (matrix-QCD$_{2,1}^{\text{adj}}$) in the 
weak coupling regime (small $g$) and in the chiral limit. The Yang-Mills potential has two distinct gauge invariant 
minima: one at $A_i=0$ and the other at $A_i = \frac{\sigma_i}{2g}$. We show that when the chiral chemical potential $c \leq \frac{3}{2}$, there is a quantum phase transition at $g_0^\ast \simeq 0.143$: for $g<g_0^\ast$, the ground state wavefunction is localized near $A_i=0$, while for $g>g_0^\ast$, the ground state is delocalized over the gauge configuration space. The transition between these two phases is singular, with the ground state at $g_0^\ast$ being distinctly different from that of $g_0^\ast \pm|\epsilon|$. At $g_0^\ast$, we show that the square of the chromoelectric field vanishes, strongly suggesting that the system is in a ``dual superconductor" phase.   Numerical evidence shows that the localization-delocalization phenomenon holds for the 1st and 2nd excited states as well, leading us to conjecture that there are an infinite number of isolated singular points $g_0^\ast> g_1^\ast>g_2^\ast> \cdots$ accumulating to $g=0$. 
For $c=1$, the model formally possesses 
$\mathcal{N}=1$ supersymmetry. We show that in the localized phase (i.e. for $g<g_0^\ast$) the supermultiplet structure is disrupted and SUSY is spontaneously broken. 
\end{abstract}

\section{Introduction} 
The gauge matrix model  \cite{Balachandran:2014iya, Balachandran:2014voa} is useful for studying the low-energy dynamics of (3+1)-dimensional QCD-like theories. Despite the drastic truncation in the number of degrees of freedom, the matrix model retains several non-trivial topological features of the full gauge theory. For instance, the gauge bundle in the matrix model is twisted \cite{Singer:1978dk, Narasimhan:1979kf} and it exhibits the axial anomaly \cite{Acharyya:2021egi}. Estimates of the glueball spectra, as well as light hadron masses using the $SU(3)$ gauge  matrix model are fairly accurate \cite{Acharyya:2016fcn, Pandey:2019dbp}. An analogous investigation of the $SU(2)$ matrix model reveals several non-trivial phases \cite{Pandey:2016hat, Acharyya:2017uhl, Acharyya:2024pqj}, both at weak and strong coupling. 

Interest in Yang-Mills quantum mechanics coupled to various kinds of quark matter is not new \cite{Witten:1981nf, Claudson:1984th, Cooper:1994eh}. A supersymmetric version of Yang-Mills quantum mechanics naturally emerges in the dimensional reduction of (9+1)-dimensional supersymmetric Yang-Mills  theory \cite{Danielsson:1996uw, Halpern:1997fv}. This direction has been pursued by practitioners of string and M-theory: it has provided insight about the dynamics of D-branes \cite{Ishibashi:1996xs, Banks:1996vh, Seiberg:1997ad, Aoki:1998bq, Berenstein:2002jq}, and is considered an ideal candidate for M-theory  \cite{Nicolai:1998ic, Dasgupta:2002hx}.

The $SU(2)$ Yang-Mills gauge theory is in a sense the simplest non-Abelian gauge theory, and provides a platform both analytical work \cite{Unsal:2007jx,Chen:2020syd,Chen:2024obv,Cordova:2018acb} and lattice-QCD simulations \cite{Astrakhantsev:2020tdl,Begun:2022bxj, Braguta:2023yhd,Iida:2024irv, Das:2025utp} to study non-perturbative properties  like the spectrum, the symmetry breaking patterns and the phase structure. 

Since the YM Matrix model has the ability to correctly capture many qualitative as well as quantitative features of (3+1)-dimensional gauge theory, it is an ideal model to study several of the above-mentioned properties. In this article, we study  the matrix model of $SU(2)$ gauge theory coupled to an adjoint Weyl fermion. This is the matrix model version of two-color one-flavor adjoint-QCD and we dub this as matrix-QCD$_{2,1}^{\text{adj}}$.

The matrix-QCD$_{2,1}^{\text{adj}}$ has $\mathcal{N}=1$ supersymmetry when the chiral chemical potential $c$ in the Hamiltonian is tuned to a specific value.  This supersymmetric model has been studied by various numerical methods \cite{Wosiek:2002nm, Campostrini:2004bs, Asplund:2015yda, Filev:2015hia, Anous:2017mwr, Asano:2018nol, Han:2019wue }, with interest in the extreme strong coupling regime.
 However, several aspects of the theory, especially the phase structure of the weak coupling regime are not well understood.

In this article, we explore the weak-coupling regime of matrix-$\text{QCD}_{2,1}^{\text{adj}}$ in the chiral limit. Our 
main tool of analysis is numerical, although we will also provide some analytic and conceptual arguments that justify 
our numerical results. The Yang-Mills potential of our model has two distinct gauge invariant minima, one at $A_i=0$ and the other at $A_i = \sigma_i/2g$, and is 
the key feature here that leads us to our results. The separation between these minima as well as the height of the intervening barrier (the saddle point) depends on $g$, as does the ground state energy. A delicate interplay between these ingredients leads to a quantum phase transition at $g_0^\ast \simeq 0.143$, as we will demonstrate. For $g<g_0^\ast$ the ground state is primarily localised at $A_i=0$, while for $g>g_0^\ast$, the ground state is delocalised over the entire available gauge configuration space. 
 Prima facie, this localization-delocalization transition appears to be different from Anderson (de)localization responsible for some types of metal-insulator transitions (for instance, see \cite{Garcia-Garcia:2006vlk,Garcia-Garcia:2005azc}).  Whether a connection emerges on further investigation is an open question.

This is perhaps an appropriate juncture to recall that phase transitions can occur even in system with finite number of degrees 
of freedom, as long as the Hilbert space is infinite dimensional \cite{Hwang:2015, Larson:2017}. Such systems can even 
demonstrate spontaneous symmetry breaking and chiral anomaly \cite{Hwang:2015,Acharyya:2021egi}. The signature of 
the phase transition is not very different from the systems with infinite  degrees of freedom: for instance, the formation 
of condensates. In the context of the quantum Rabi model, it is well-known that there is a phase transition controlled by 
the coupling: one phase being normal and the other being superradiant. At the phase transition, the ground state  becomes translationally invariant signalling the formation of a condensate \cite{Hwang:2015}.

Representations of the 
Heisenberg-Weyl algebra with a translationally invariant ground state are non-regular \cite{Beaume:1974ad, Acerbi_combo}, and are not 
unitarily equivalent to the regular representation (see appendix \ref{nonregular_rep_HW_algebra}).  The non-regular representation has some unusual but well-understood properties.  For instance,  the expectation value of 
 the positive operator $\widehat{x}^2$   in this state is infinite. 
A mathematically rigorous way of saying this is that $\widehat{x}^2$  is not an observable.  For the matrix 
model, although the details are different from  the quantum Rabi model, the situation is conceptually identical. 

The inequivalence betweeen regular and non-regular representations has rather startling consequences for our discussion here. 
As the Yang-Mills coupling $g \rightarrow g_0^\ast$, the limiting ground state wavefunction becomes non-normalizable in the Hilbert space. The ground state at $g_0^\ast$ has a fundamentally different character compared to that for $g \neq g_0^\ast$. The loss of normalizability of the limiting wavefunction is due to its non-trivial support at the saddle point, which in turn can be thought of as the formation of a condensate.

This phenomenon is not limited to just the ground state, but also happens for eigenstates with higher energies at their own critical points,  leading to the existence of phase transitions at higher energies as well.

We start our investigation with the chiral chemical potential $c$ set to zero and establish this quantum phase transition. Turning on $c \neq 0$ introduces an additional level crossing at $g_0^R(c)$, where the fermion number of the ground state changes. The localization-delocalization transition at $g_0^\ast$ can then only occur if  $g_0^R(c) < g_0^\ast$. We find that this is possible only as long as $c\leq \frac{3}{2}$.  

The model has a formal $\mathcal{N}=1$ supersymmetry  for $c=1$, which is directly affected by this localization phenomenon. In the localized phase ($g < g_0^\ast$), the pairing between bosonic states within a supermultiplet is disrupted, which in turn breaks supersymmetry for all $g< g_0^\ast$.

This article is organized as follows. In Section 2, we discuss the theoretical framework of matrix-$\text{QCD}_{2,1}^{\text{adj}}$ and argue for the existence of localization-delocalization transition at $g_0^\ast$. In Section 3, we outline our numerical strategy to estimate the energy eigenvalues and eigenfunctions. Following this, we present the numerical evidence for the localization-delocalization transition in Section 4. In the same section, we discuss the effect of the chiral chemical potential and the fate of $\mathcal{N}=1$ SUSY. We end with a discussion in Section 5.

\section{Matrix-$\text{QCD}_{2,1}^{\text{adj}}$: Theoretical considerations }\label{sec_2}
 The gauge degrees of freedom (the glue) are described by a real matrix $M_{ia}$ (with spatial index $i=1,2,3$ and color index $a=1,2,3$), or alternately by $A_{i} \equiv M_{ia} T^a$, with $T^a$'s being the generators of $SU(2)$ in the fundamental representation. The phase space for the pure gauge theory is  generated by $M_{ia}$ and their conjugate momenta $P_{ia} \equiv - i \frac{\partial \,}{\partial M_{ia}} $.  The chromoelectric field $\Pi_{i} \equiv P_{ia} T^a$  and the chromomagnetic field $B_{i} \equiv -\rho^{-1} A_i - \frac{i}{2} \epsilon_{ijk} [A_j, A_k]$ give us the pure Yang-Mills Hamiltonian 
 \begin{eqnarray}
 H_{YM} =\text{Tr} \Big( \frac{g^2}{2 \rho^3} \Pi_i \Pi_i +  \frac{\rho^3}{2g^2} B_i B_i \Big)
 \end{eqnarray}
 where $\rho$ is radius of the spatial $S^3$ and $g$ is the Yang-Mills coupling.

The adjoint Weyl quark $b_{\alpha a }$ ($\alpha=1,2$,  $a=1,2,3$) is a time-dependent Grassmann-valued matrix \cite{Pandey:2016hat}, interacting with the glue via minimal coupling:  $H_{int}= 2  i \epsilon_{abc}\rho^{3}  b_{\alpha b}^\dagger  \sigma^{i}_{\alpha\beta} b_{\beta c} \text{Tr} (T^a A_i)$. We also add a chiral chemical potential term $H_c=c \rho^{2}  b_{\alpha b}^\dagger b_{\alpha c}$ \cite{Acharyya:2024pqj}, and investigate the system's dependence on $c$, and in particular, its supersymmetric properties. Finally, one may add a quark  mass term, but we will consider only the  chiral limit   (i.e. massless case) here.

Dimensionless variables are  convenient to work with, so we rescale $A_i \to g^{-1}\rho A_i$, $\Pi_i \to g \rho^{-1} \Pi_i$ and $b_{\alpha a} \to \rho^{\frac{3}{2}} b_{\alpha a}$.  In terms of  the rescaled variables, the Hamiltonian is $H = H_{YM} + H_{int}+ H_c$: 
  \begin{eqnarray}
 H =\frac{1}{\rho}  \text{Tr} \Big( \Pi_i \Pi_i + A_i A_i + i g \epsilon_{ijk} [A_i, A_j]A_k -\frac{g^2}{2} [A_i, A_j]^2 +  2i g \epsilon_{abc} b_{\alpha b}^\dagger  \sigma^{i}_{\alpha\beta} b_{\beta c}  T^a A_i + c \,Q_R  \Big) \label{Ham_1}
 \end{eqnarray}
where $Q_R  \equiv (b^\dagger_{\alpha a} b_{\alpha a}-3)$  and   $\rho^{-1}$ determines the energy scale. In our investigation, we will fix $\rho=1$ and study the theory as a function of the coupling $g$. This is equivalent to fixing $g$ and varying $\rho$  or equivalently, the volume \cite{Acharyya:2024pqj}.

The glue Hilbert space $\mathcal{H}_G=\displaystyle{L^2(M_3(\mathbb{R}),\prod_{ia} dM_{ia})}$ is infinite dimensional and the glue states may be organized in representations of spin and color.  The quark Hilbert space $\mathcal{H}_F$ is  64-dimensional, and the fermionic states can have integer or half-integer spin $0\leq s \leq 3/2$ and   transform in singlet, triplet or quintuplet representations of color $SU(2)$. These, and other symmetries of the system are  recalled in appendix~\ref{app_sym}. 

The total Hilbert space is $\mathcal{H}_G \otimes \mathcal{H}_F$, where  the states may also be labelled by spin and color.  The physical Hilbert $\mathcal{H}_{phys} \subset \mathcal{H}_G \otimes \mathcal{H}_F$ is the set of colorless states (i.e those annihilated by the Gauss law (\ref{gauss_law1},\ref{gauss_law2}): 
\begin{eqnarray}
G_a | \phi \rangle =0, \quad \quad [G_a, G_b]=i \epsilon_{abc} G_c, \quad  | \phi \rangle \in \mathcal{H}_{phys}. 
\end{eqnarray}

\begin{figure}
\begin{center}
\includegraphics[width=18cm]{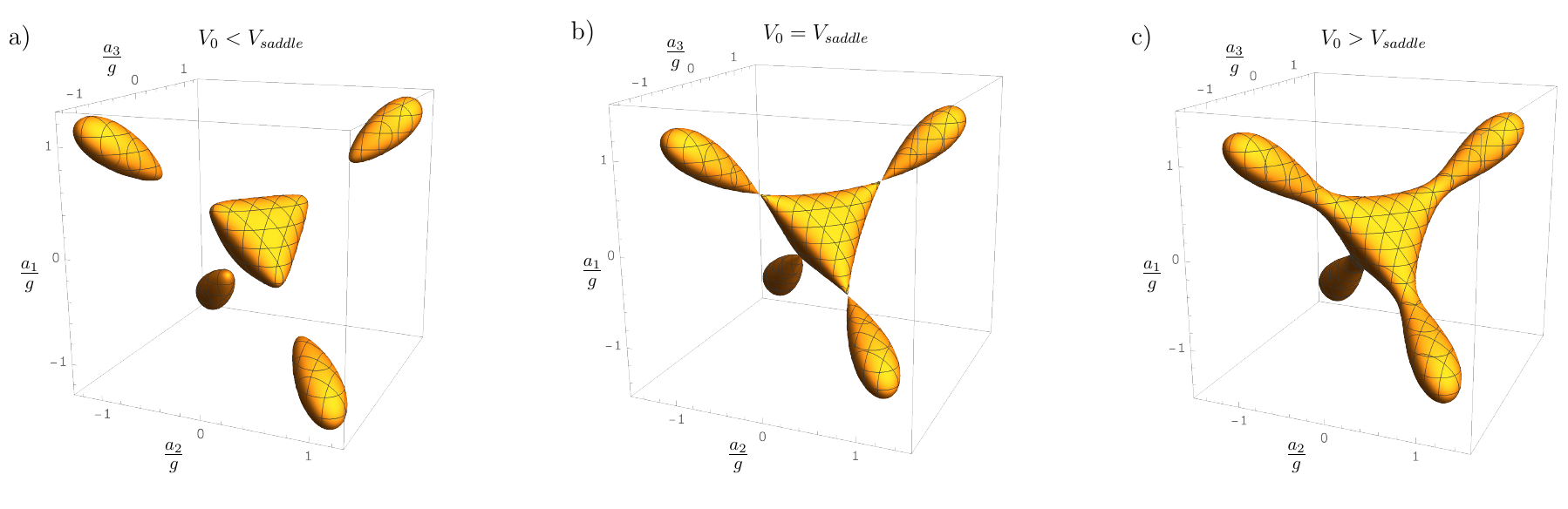}
\caption{ Equipotential surfaces of $V_{YM}$ in the $a_1$-$a_2$-$a_3$ space.  
}\label{Fig_equipot} 
\end{center}
\end{figure}

 A discussion of the extrema of the Yang-Mills potential provides important qualitative information. Recall that 
 \begin{eqnarray}
 V_{YM} = \text{Tr} \Big( A_i A_i + i g \epsilon_{ijk} [A_i, A_j]A_k -\frac{g^2}{2} [A_i, A_j]^2\Big)
 \end{eqnarray}
is minimized when the $A_i$ obey $[A_i, A_j]=\frac{i}{g}\epsilon_{ijk}A_k$. This has two solutions, $A_i=0$ and $\frac{1}{2g}\sigma_i$.  There is also an extremum at  $A_i= \frac{1}{4g}\sigma_i $. Evaluating the Hessian  at this extremum shows that it is a saddle point.  This saddle point is the matrix model analogue of the sphaleron -- a static solution of the pure YM equation with just  one unstable direction \cite{Asorey:1996klg}.

\begin{figure}[t!]
\begin{center}
\includegraphics[width=18cm]{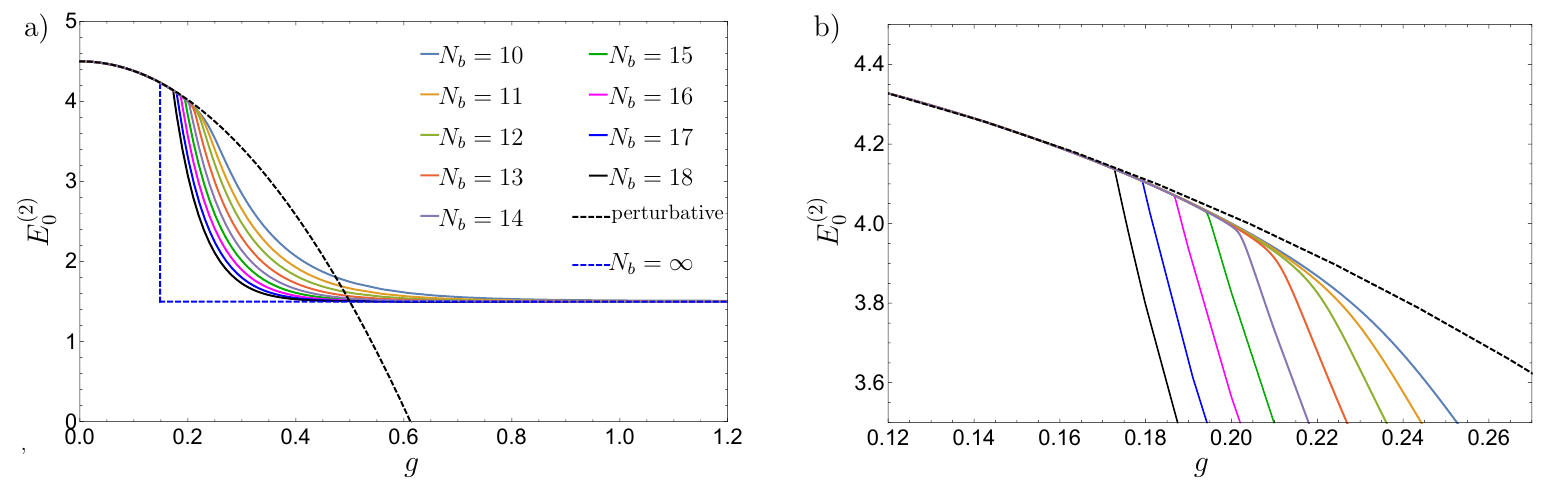}
\caption{a) $E_0^{(2)}$ as a function of $g$ for various  $N_{b}$.  The solid lines represent the data from Rayleigh-Ritz method.  The black dashed lines shows the estimate from perturbation theory as in Eq.(\ref{pert_gs_E}). The blue dashed line is the extrapolation to the $N_{b} \to \infty$ limit.  b) The region around the kink in $E_0^{(2)}(g)$.  The colored curves here  follow the same labelling as in Fig.~3a.
Here  $c=0$ for both the figures.}\label{Fig_1} 
\end{center}
\end{figure}

At the minima, the potential $V_{YM}=0$, while at the saddle  $V_{YM}=\frac{3}{32g^2}$. Further insight is obtained by looking at the frequencies $\omega$ of small oscillations at each minimum (see appendix~\ref{app_min_rect} for details). Around $A_i=0$ we find that $\omega=1$ with nine-fold degeneracy, while near $A_i=\frac{1}{2g}\sigma_i$ we have $\omega=2$ (5-fold degenerate) and a non-degenerate $\omega=1$. There are three flat directions as well, but they correspond to  gauge transformations. On quantizing these modes, the zero-point energy near $A_i=0$ is $\frac{9}{2}$, while for near $A_i=\frac{1}{2g}\sigma_i$ it is $\frac{11}{2}$. Hence despite the structure of the potential being that of a double well,  the ground state for very small $g$ preferably localizes near $A_i=0$ and is therefore unique. As we will show in the subsequent sections, this heuristic argument is supported by the numerical analysis.  As we increase $g$, we find that the quark-glue interaction reduces  the energy of the ground state (see Fig.~\ref{Fig_1}a).  But increasing $g$ also reduces the barrier height which can lead to a ground state delocalized over both the wells.  Remarkably, we find that this ``localization-delocalization'' transition happens sharply at a small  but finite $g$.

Coordinates of the singular value decomposition (SVD), rather than the rectangular coordinates $M_{ia}$, provide valuable qualitative information as well. In these coordinates, the rotations and gauge transformations can be neatly separated, and the description of the extrema of the potential becomes more transparent. This separation also helps in revealing certain crucial properties of the Hamiltonian, like its domain of definition in $\mathcal{H}_{phys}$, which otherwise remains hidden when described in the rectangular coordinates, as we argue below. 

Recall that a real matrix $M_{ia}$ can be decomposed as 
\begin{eqnarray}
M = R D S^{T}, \quad R \in O(3)_{rot}, \quad S \in Ad(SU(2))\simeq SO(3)_{col}, \quad D = \text{diag}(a_1, a_2, a_3), \quad a_i \in \mathbb{R}. 
\end{eqnarray}
The $a_i$'s are the singular values of $M$ and using $O(3)_{rot}$ rather than $SO(3)_{rot}$ allows us to include parity. The $a_i$'s are both rotation- and gauge-invariant. The kinetic energy in the SVD coordinates was first presented in \cite{Iwai:2010}. 
The integration measure is 
$(\sqrt{\phi} da_1 da_2 da_3 d\Omega_{rot} d\Omega_{col})$ where $\sqrt{\phi} = |(a_1^2-a_2^2)(a_1^2-a_3^2)(a_2^2-a_3^2)|$, and  $d\Omega_{rot}$ and  $d\Omega_{col}$ are the invariant volume forms on $SO(3)_{rot}$ and $SO(3)_{col}$ respectively. 
By a scaling transformation, we can get rid of the awkward $\sqrt{\phi}$ factor in the measure, allowing us to work with the flat measure $da_1 da_2 da_3$ in the $a_i$-directions \cite{Acharyya:2017uhl}.

After this scaling transformation, the kinetic and potential energy terms in the Hamiltonian become 
\begin{eqnarray}
\text{Tr } \Pi_i \Pi_i &=& \sum_{i}\Big[ -\frac{1}{2}\frac{\partial^2}{\partial a_i^2}+ \frac{1}{4} \sum_{j\neq i}\frac{a_i^2 + a_j^2}{(a_i^2- a_j^2)^2}+ \frac{1}{4} \sum_{j\neq i, k \neq j, i} \frac{(a_j^2 + a_k^2) (\widetilde{L}_{i}^2+\widetilde{G}_{i}^2) + 4 a_j a_k \widetilde{L}_{i } \widetilde{G}_{i} }{(a_j^2 - a_k^2)^2} 
\Big], \label{kin_sing} \\
V_{YM} &=& \left( \frac{1}{2} (a_1^2 + a_2^2 + a_3^2) - 3 g a_1 a_2 a_3 + \frac{g^2}{2} (a_1^2 a_2^2 +a_2^2 a_3^2+a_3^2 a_1^2)\right).  \label{pot_sing} 
\end{eqnarray}
Here $ \widetilde{L}_{i }$ and $\widetilde{G}_{a}$ are generators of ``body-fixed'' spatial and gauge rotations respectively, satisfying $  [ \widetilde{L}_{i },  \widetilde{L}_{j}]=-i\epsilon_{ijk}  \widetilde{L}_{k }$ and $[\widetilde{G}_{a},\widetilde{G}_{b}]=-i\epsilon_{abc} \widetilde{G}_{c}$.

A short computation from (\ref{pot_sing}) shows that the minima are at $a_i=0$ and $a_i=1/g$, while the saddle point is at $a_i=1/2g$. The equipotential surfaces of $V_{YM}$ are shown in Fig. \ref{Fig_equipot}. When $V_{YM}< \frac{3}{32g^2}$, the equipotential surfaces consists of five disconnected lobes (Fig.\ref{Fig_equipot}a), while for $V_{YM} > \frac{3}{32g^2}$ the lobes are connected (Fig.\ref{Fig_equipot}c). The value of $V_{YM}=\frac{3}{32g^2}$ is special, and the connection of the lobes happen at isolated points in $a_1$-$a_2$-$a_3$ space, precisely at the saddle point (Fig.\ref{Fig_equipot}b).

In the SVD coordinates, the system has invariance under certain discrete transformations. The potential $V_{YM}$ is invariant under permutations $S_3$ i.e. $a_i \to a_{\sigma_{(i)}}$. It is also invariant under the Klein group $\mathbb{Z}_2 \times \mathbb{Z}_2$ with $a_i \to \epsilon_ia_i$  ($\epsilon_i = \pm 1$, $\epsilon_1 \epsilon_2 \epsilon_3 =1$). The $S_3$ and the $\mathbb{Z}_2 \times \mathbb{Z}_2$ can be combined into the tetrahedral group $T$, and it may seem that the symmetry of the Hamiltonian is $T$. But this is not quite correct, since the $\mathbb{Z}_2 \times \mathbb{Z}_2$ are gauge transformations, and the set of gauge-inequivalent $a_i$'s is $D/(\mathbb{Z}_2 \times \mathbb{Z}_2)$. The quotient $D/(\mathbb{Z}_2 \times \mathbb{Z}_2)$ is a little awkward to visualize, and we will work with $D$ for drawing pictures. This is harmless as long as we remember at the physical configuration space is quotiented by the $\mathbb{Z}_2 \times \mathbb{Z}_2$. The kinetic operator (\ref{kin_sing}) obviously possesses the same invariance along with the same qualifications, provided the components of $ \widetilde{L}_{i }$ (and $\widetilde{G}_{a}$) are also exchanged appropriately.

The kinetic term (\ref{kin_sing}) appears to be singular whenever $|a_i|=|a_j|$. This may seem puzzling, since in the rectangular coordinates, the kinetic term is free of any singularities. 
This simply means that the domain of the kinetic operator consists of normalizable states in  $\mathcal{H}_{phys}$ that vanish on the surface $|a_i|=|a_j|\neq 0$.  All such wavefunctions have nodes (i.e. vanish) at $|a_i|=|a_j|$. Physically, this is consistent with the fact that the gauge bundle  $M_3(\mathbb{R})/Ad(SU(2))$ is twisted. 

It is important to note that the saddle point lies on the surface $|a_i|=|a_j|$. Therefore all wavefunctions in the domain of the kinetic operator vanish at the saddle point, and hence the probability of finding the particle here is zero. Conversely functions of $a_i$ that do not vanish at the saddle point are not in the domain of the kinetic  operator.

For small $g$, which is the regime we are interested in, we can use perturbation theory to make some estimates to decide on the importance of the saddle point. When $g$ is small, the two minima are widely separated and the barrier height between them is very large. As we argued earlier, a particle with ground state energy less than the barrier height $\frac{3}{32g^2}$ will preferentially localize at $a_i=0$. On the other hand, if the energy of the particle is greater than  $\frac{3}{32g^2}$, the particle does not remain localized around the central minimum and travels to the regions near the other minimum as well. As $g$ is increased, the minima move closer and the barrier height between them decreases.  When the energy is exactly equal to $\frac{3}{32g^2}$, there is a new (unstable) classical solution -- the sphaleron -- corresponding to the particle standing still at the top of the barrier.  This situation is quantum mechanically forbidden as the wavefunction must vanish on the saddle point.

In the following section, we show that  in presence of the quark-glue interaction, the dynamics of the quantum system in the weak coupling regime reproduces this classical picture to a large extent. 
Our main result, a rather striking one, is that as the coupling approaches a particular value $g_{0}^{\ast}$, the ground state gets localized on the surface $|a_i|=|a_j|\neq 0$. Using finite size scaling, we will show that in fact at $g_{0}^{\ast}$, the putative ground state, because of  its localization on this surface, is no longer square-integrable and hence not a member of $\mathcal{H}_{phys}$.

To summarize, for $g<g_0^\ast$, the ground state remains localized near the minimum at $a_i=0$, while for $g>g_0^\ast$, it describes the delocalized phase. At the interface of these two phases,  the point $g_0^\ast$ itself corresponds to another distinct phase. We will argue below that this phase at $g_0^\ast$ corresponds to the formation of a condensate and has properties like a dual superconductor.

In fact, we will show that the 1st as well as the 2nd excited states  with non-zero fermion number $n_F$ behave in this way, with their critical couplings $g_1^\ast$ and $g_2^{\ast}$ obeying $g_2^\ast < g_{1}^{\ast} < g_{0}^{\ast}$. This leads us to conjecture that the $n^{\text{th}}$ excited state also behaves similarly, with its critical  coupling  $g_{n}^{\ast} < g_{n-1}^{\ast}$.  As the spectrum is discrete, the critical points $g_{n}^{\ast}$ are isolated.  At every $g_{n}^{\ast}$, one state acquires non-zero support at the saddle point and 
hence becomes non-normalizable. 

When $c=1$ the Hamiltonian (\ref{Ham_1}) formally has ${\cal N}=1$ supersymmetry (see appendix~\ref{app_sym}). 
What happens to SUSY at these critical points?  The SUSY multiplets rely on a delicate balance between bosonic and fermionic states. As we will show,  the equality between the number of bosonic and fermionic states is violated for every interval $g_{n+1}^{\ast}<g<g_{n}^{\ast}$.  As a consequence, SUSY is broken for all $g<g_{0}^{\ast}$.

\section{Strategy for Variational Estimation of Energies} 
Our particular interest is in the color-singlet energy eigenstates. Since the total spin commutes with the Hamiltonian, these states can in addition be labelled by $J$, which is an integer or half-integer. Fermion number is also a useful label as it is conserved almost for all values of $g$ (it is ill-defined at the locations of level crossings).

\begin{figure}
\begin{center}
\includegraphics[width=18cm]{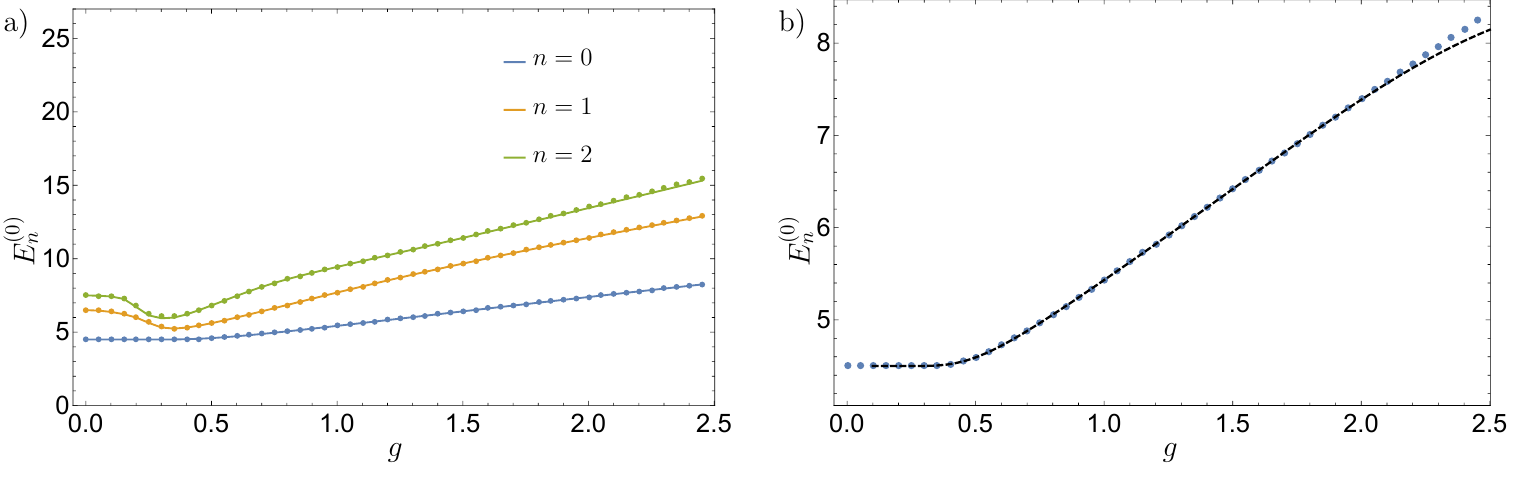} 
\caption{ a) The energies of the states in $n_F=0$ sector: $E_n^{(0)}$ as function of $g$ for $n=0,1,2$.  The dots represent the data with $N_{b}=18$ and the solid lines with $N_{b}=16$. b) The comparison of $E_n^{(0)}$ with the corresponding fit in (\ref{fit_E0_nf0}), which is represented by the black dashed line.  Here, we have chosen $c=0$. }\label{Fig_En_nf0_sector} 
\end{center}
\end{figure}

The colorless states are composites $|\psi_{glue}\rangle \otimes |\phi_{quark}\rangle$, where $|\psi_{glue}\rangle$ are expanded in the basis of eigenstates of the 9-dimensional harmonic oscillator.   We consider a finite-dimensional trial set for $|\psi_{glue}\rangle$ the by setting a  boson number cutoff $N_b$. 
Trial wavefunctions are constructed by taking arbitrary linear combinations of states from the trial set. By varying the coefficients of this expansion, the variational method discovers the specific state that minimizes the energy. Expectation values of various interesting observables can now be computed in the minimum energy eigenstate.  These states belong to the $\mathcal{H}_{phys}$, and we explicitly verify that these states vanish on the surface $|a_i|=|a_j|>0$ (see appendix~\ref{app_SVD_1}).

In general, any numerical investigation of a physical system involves a cutoff (like our  $N_{b}$, or inverse lattice spacing $\Lambda$ in  the case of lattice QCD). For generic values of the parameters of the  Hamiltonian, expectation values of observables converge to their true values as $N_{b} \to \infty$. However, this may not be true for parameter values near a phase transition. 
Even though there is loss of convergence, we can use finite-size scaling (FSS) to determine the dependance of  the expectation value on $N_{b}$. From here we can deduce the behavior of that expecation value as $N_{b} \to \infty$.

We progressively  increase starting from $N_{b}=10$  till we reach convergence in energy and other observables. We have obtained data till $N_{b} =18$, though for generic values of $g$, we have found that $N_{b}=16$ is sufficient to achieve excellent convergence.

Near $g_0^\ast$, we find that the ground state expectation values of observables do not converge. We will argue that this is a signal of the ground state becoming non-normalizable at $g_0^\ast$. 
To study this phenomenon in a controlled manner, we will use FSS in $N_{b}$ and hence we will be able to make precise statements about the observables at $g_0^\ast$. Similar remarks hold in the neighbourhood of each $g_n^\ast$ as well. Here of course, it is the $n^{\text{th}}$ excited state that becomes non-normalizable.

\section{Results} 
\subsection{Low-lying energy eigenstates in absence of chiral chemical potential} 

\begin{figure}[b!]
\begin{center}
\includegraphics[width=8cm]{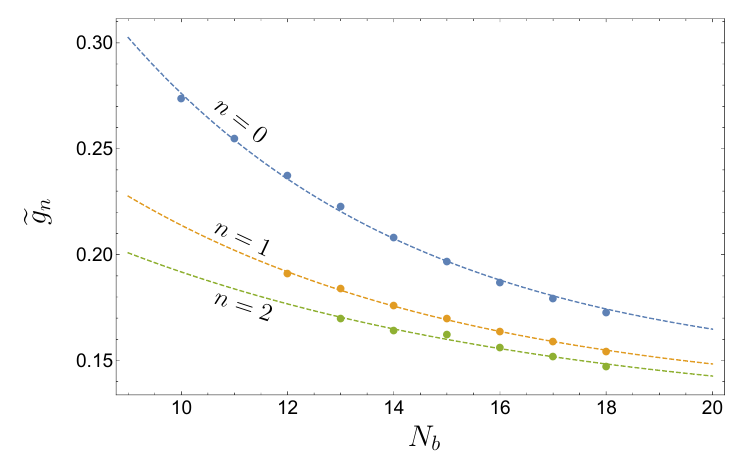}
\caption{ $\widetilde{g}_n$ vs $N_{b} $ for $n=0,1,2$.  The dots  represent the numerical data using Rayleigh-Ritz method and the dashed lines are  fits
in (\ref{fit_g0}) and  (\ref{fit_gn}) with parameters in Table \ref{Table-2}.  Here, we have chosen $c=0$.  }\label{Fig_gnstar} 
\end{center}
\end{figure}
Using the numerical strategy described above, we obtain the low-lying spin-0 energy eigenstates. For $c\geq 0$, 
these states have $n_F=0$ or $n_F=2$. 
In this section, we set  $c$ to zero and label the energies of these states as  $E_{n}^{(n_F)}(g)$.

The energies $E_{n}^{(0)}(g)$ of the three lightest states in the $n_F=0$ sector (as functions of  $g$) are shown in Fig.~\ref{Fig_En_nf0_sector}a. These energy eigenvalues exhibit excellent convergence with $N_{b}\geq 16$ in the $g \in [0,2.5]$ range. Further, we find that the energy of the lightest state with $n_F=0$ for small $g$ can be fitted to
\begin{eqnarray}
E_{0}^{(0)} \approx \frac{9}{2} + (1.053 + 0.775 g^2  -0.047 g^4+ \ldots)e^{- 0.647/g^2}. \label{fit_E0_nf0}
\end{eqnarray}
The comparison of the above with the numerical data is shown in Fig.~\ref{Fig_En_nf0_sector}b.
 \begin{table}
\begin{center}
\begin{tabular}{|c||c|c|c|c||c|c|c|c|c}
\hline 
State& $n$ &$a_n$ & $b_n$   & $g_{n}^{\ast}$ & $g_{n}^{\ast}$ & $d_n$ & $k_n$  \\ 
& & & &(from perturbation & (from numerical && \\ 
& & & &theory) & data) && \\ \hline  \hline  & & & & &&&\\ 
$|\Psi_{0}^{(2)}\rangle $ & 0 & 4.5  & -12 & 0.1488 & 0.143 & 0.81 & 0.181 \\ 
& & & & &&& \\ 
$|\Psi_{1}^{(2)}\rangle $ & 1 & 5.5  & -6.72 & 0.1320 & 0.13 & 0.382 & 0.152  \\ 
& & & & &&& \\ 
$|\Psi_{2}^{(2)}\rangle $ & 2 & 6.5  & -30.12 & 0.1247 & 0.123 & 0.236 & 0.122  \\ 
  \hline 
\end{tabular}
\caption{Parameters for the fits of the $E_{n}^{(2)} \approx a_n + b_n g^2$, the estimated values ${g}_{n}^{\ast}$ from perturbation theory using (\ref{En_eq_saddle}), and from numerical data using the fits of the $\widetilde{g}_n$ in (\ref{fit_g0}) and (\ref{fit_gn}). }\label{Table-1}
\end{center}
\end{table}

The ground state has $n_F=2$ for all $g>0$ (compare Fig.~\ref{Fig_En_nf0_sector}a and Fig.~\ref{Fig_1}a).  We will study the 
following gauge- and rotationally-invariant observables and their expectation values : 
\begin{eqnarray}
&& \Delta \equiv(-\partial H/\partial g)=  - \text{Tr} \Big( i \epsilon_{ijk} [A_i, A_j]A_k -2g [A_i, A_j]^2 +  2i \epsilon_{abc} b_{\alpha b}^\dagger  \sigma^{i}_{\alpha\beta} b_{\beta c}  T^a A_i\Big), \nonumber  \\
&& \Phi \equiv 2\text{Tr }  A_{i} A_i, \quad\quad K \equiv \text{Tr} \left( A_i A_i + i \frac{g}{3} \epsilon_{ijk}[ A_i, A_j] A_k \right) \\ 
&& \mathcal{O}_{1, n} \equiv -\frac{\partial E_{n}^{(2)}}{\partial g} =\langle \Psi_n^{(2)} |\Delta |\Psi_{n}^{(2)} \rangle, \quad\quad   \mathcal{O}_{2,n} \equiv \langle \Psi_n^{(2)} | \Phi |  \Psi_n^{(2)} \rangle, \quad\quad \mathcal{O}_{3,n} \equiv \langle \Psi_n^{(2)} | K |\Psi_{n}^{(2)} \rangle. \nonumber 
\end{eqnarray}
Here, $K$ is our analogue of the Hodge-dual of the Chern-Simon 3-form \cite{Witten:1982df} satisfying $[K, \Pi_i] =- i B_i$ and $[K, A_i]=0$. 
In addition to  the energies of the states, we will also track the $ \mathcal{O}$'s as functions of $g$.  

Since  $\Delta$, $\Phi$ and $K$ as operators are polynomials in $A_i$, their expectation values  in any physical state should be non-singular functions of $g$. 
But what are we to make of the physics if their expectation values  grow in an unbounded manner at some finite $g$? One possible reason why this may  happen is if the respective domains of the observable and $H$ are different. This is for instance what happens when there is an anomaly \cite{Esteve:1986db, Balachandran:2011bv}.  A second possibility, which happens to be the case at hand, is that the state used to compute the expectation value becomes non-normalizable at that $g$.  This is the signal of the formation of a non-trivial condensate. 

\begin{figure}[b!]
\begin{center}
\includegraphics[width=18cm]{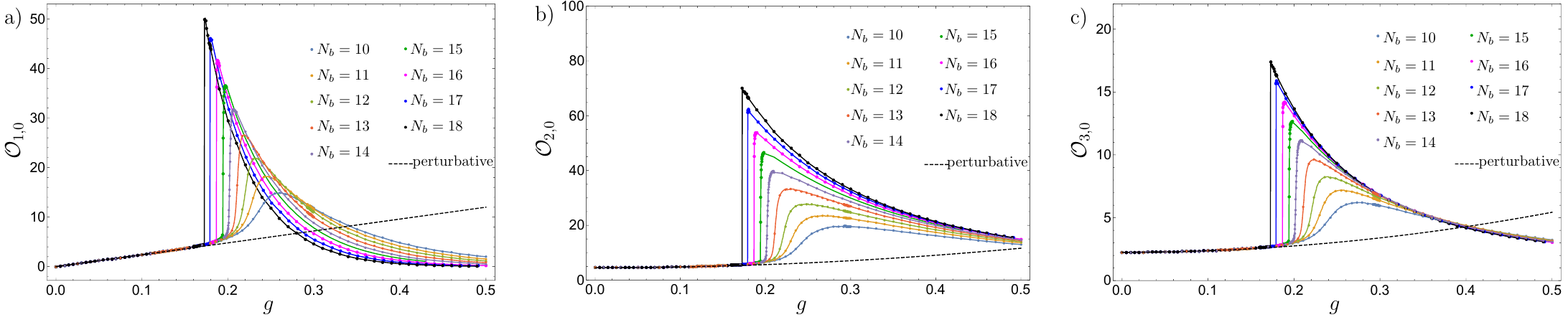} 
\caption{ The expectation values $\mathcal{O}_{A,0}$ as a function of $g$ for various  $N_{b}$ in the $0\leq g \leq 0.5$ regime.  Black dashed lines corresponds to the estimate from perturbation theory:
$   \mathcal{O}_{1,0} \simeq 24 g + O(g^3)$,  $\mathcal{O}_{2,0} \simeq \frac{9}{2} + \frac{57}{2} g^2 + O(g^4)$, and $ \mathcal{O}_{3,0} \simeq \frac{9}{4} + \frac{51}{4} g^2 + O(g^4)$.   Here, we have chosen $c=0$. }\label{Fig_3} 
\end{center}
\end{figure}

To begin with, we numerically diagonalize the Hamiltonian in the $n_F=2$ sector to obtain its ground state  $|\Psi_0^{(2)}\rangle$ and energy for $g\in [0,1.5]$ (Fig.~\ref{Fig_1}a).  At small $g$,  there is a kink in the ground state energy. The location $\widetilde{g}_0$ of the kink depends on $N_{b}$ (Fig.~\ref{Fig_gnstar}), demonstrating loss of convergence near $\widetilde{g}_0$.   Away from the kink, there is excellent convergence on either side. 
Finite-size scaling gives us
\begin{eqnarray}
\widetilde{g}_0 (N_{b}) =g_{0}^{\ast}   +  d_0 \, e^{-k_0 N_{b}}, \quad \text{where}\quad g_{0}^{\ast} \simeq 0.143, \quad d_{0}\simeq 0.81, \quad k_{0}\simeq 0.181,  \label{fit_g0}
\end{eqnarray}
thus identifying the true location $g_{0}^{\ast}$ of the kink.

\begin{figure}
\begin{center}
\includegraphics[width=16cm]{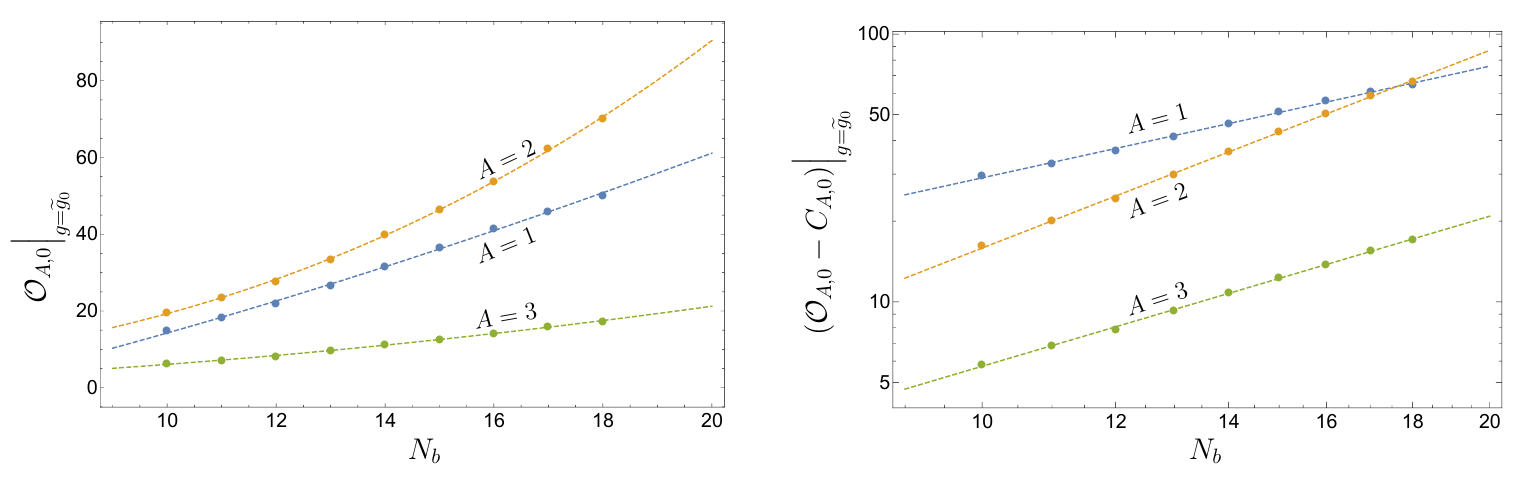}
\caption{Log-log plot of $\mathcal{O}_{A,0}\Big|_{g=\widetilde{g}_0}$ vs $N_{b} $.  The dots  represent the numerical data using Rayleigh-Ritz method and the dashed lines are  fits
in (\ref{power_law_1})  with parameters in Table \ref{Table-2}.  Here, we have chosen $c=0$.  }\label{Fig_5} 
\end{center}
\end{figure}

The situation is even more dramatic for the observables $\Delta$, $\Phi$ and $K$, as these show  power-law divergences (Fig.\ref{Fig_3}) when we approach $g_{0}^{\ast}$ from the right:
\begin{eqnarray}
 \mathcal{O}_{A,0}\Big|_{g=\widetilde{g}_0}= C_{A,0}+  \alpha_{A,0} (N_{b})^{\beta_{A, 0}}. \label{power_law_1}
\end{eqnarray}
The parameters  $C_{A,0}$, $\alpha_{A,0}$ and   $\beta_{A, 0}$ are given in Table \ref{Table-2}, and the power-law behavior is  confirmed by the straight line fit of the log-log plot in  Fig.~\ref{Fig_5}.    

\begin{figure}[b!]
\begin{center}
\includegraphics[width=18cm]{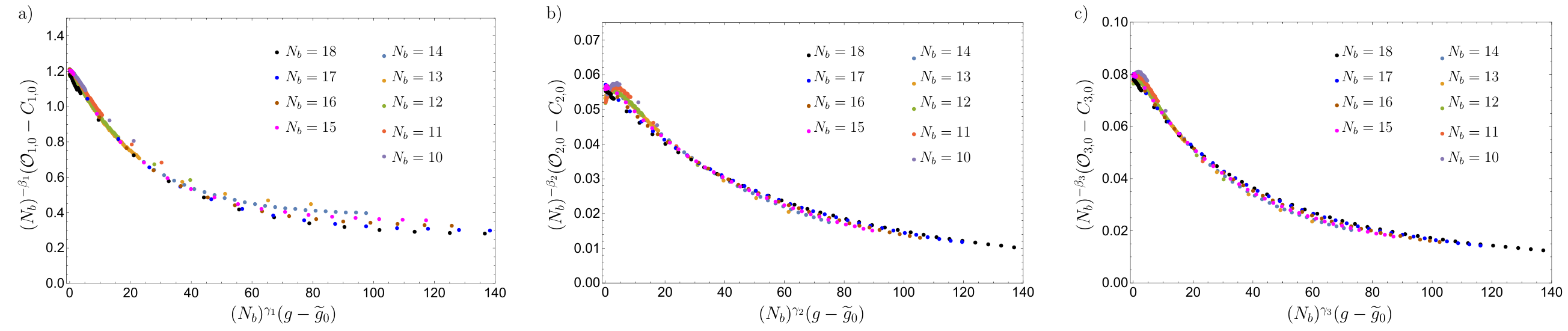}
\caption{$(N_{b})^{-\beta_{A,0} }(\mathcal{O}_{A,0}-C_{A,0})$ as a function of $(N_{b})^{\gamma_{A,0} } (g-\widetilde{g}_0)$ for various  $N_{b}$. The parameters $C_{A,0}$, $\beta_{A,0}$and $\gamma_{A,0}$ are given in Table \ref{Table-2}.  }\label{Fig_6} 
\end{center}
\end{figure}

Further investigation reveals that $(N_{b})^{-\beta_{A,0}}  (\mathcal{O}_{A,0}-C_{A,0}) $ as function of $(N_{b})^{\gamma_{A,0}}(g-\widetilde{g}_0)$  is independent of $N_{b}$ for all $g>\widetilde{g}_0(N_{b})$ (Fig.~\ref{Fig_6}), with $\gamma_{A,0}$  given in Table~\ref{Table-2} (for a discussion of FSS in quantum mechanical systems, see for example \cite{Neirotti:1998}).  
Our data collapse strongly suggests that the region of non-convergence shrinks to just the critical point at $g=g_{0}^{\ast}$ as $N_{b} \to \infty$.

\begin{figure}[t!]
\begin{center}
\includegraphics[width=15cm]{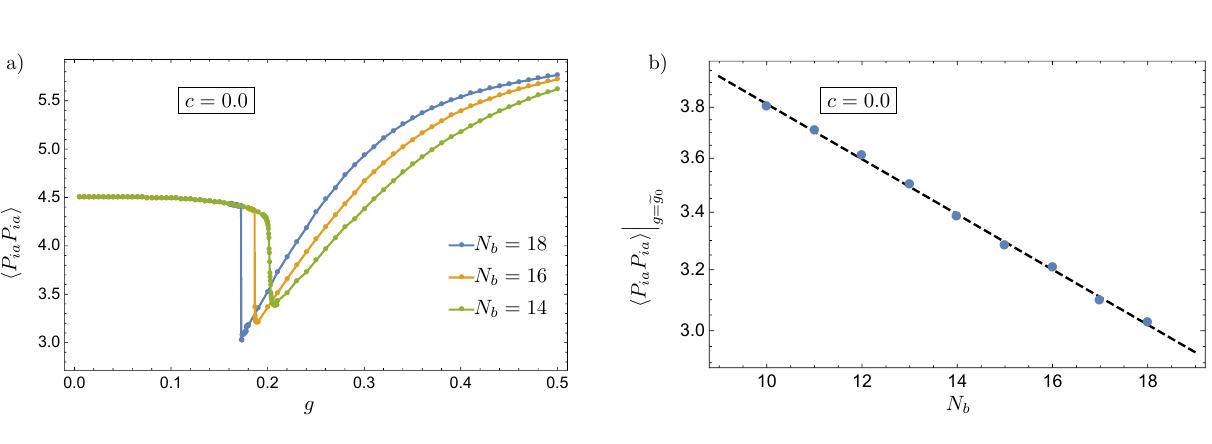}
\caption{a) The expectation value of  chromoelectric field squared $\langle P_{ia} P_{ia}\rangle$ in the state $|\Psi_0^{(2)}\rangle$ as a function of $g$ for different $N_b$. b) The linear behavior of $\log\langle P_{ia} P_{ia}\rangle$ at $g=\widetilde{g}_0$ as a function of $N_b$: $\langle P_{ia} P_{ia}\rangle \approx 5.1 e^{-0.03 N_b}$ (black dashed line). This implies that  $\langle P_{ia} P_{ia}\rangle$ at $g=\widetilde{g}_0$ vanishes as $N_b \to \infty$. }\label{Fig_electric field} 
\end{center}
\end{figure}

The scaling behavior of $\Delta$ convincingly shows that there is a step-function discontinuity in $E_0^{(2)}$ at $g=g_0^\ast$.  This in turn implies that the putative ground state in the FSS limit is no longer normalizable at $g_0^\ast$. As we will show below, the 1st excited state is well-behaved  at $g_0^\ast$ and is the true ground state at this coupling. It is precisely in this sense that there is a QPT at $g_0^\ast$.

We can deduce some other interesting properties of the system from $\Phi$ (Fig.~\ref{Fig_3}b). It remains rather small for $g<g_0^\ast$, implying a localization to the region of small $A_i$.  It becomes significantly large just above $g_0^\ast$.  

\begin{figure}[t]
\begin{center}
\includegraphics[width=17cm]{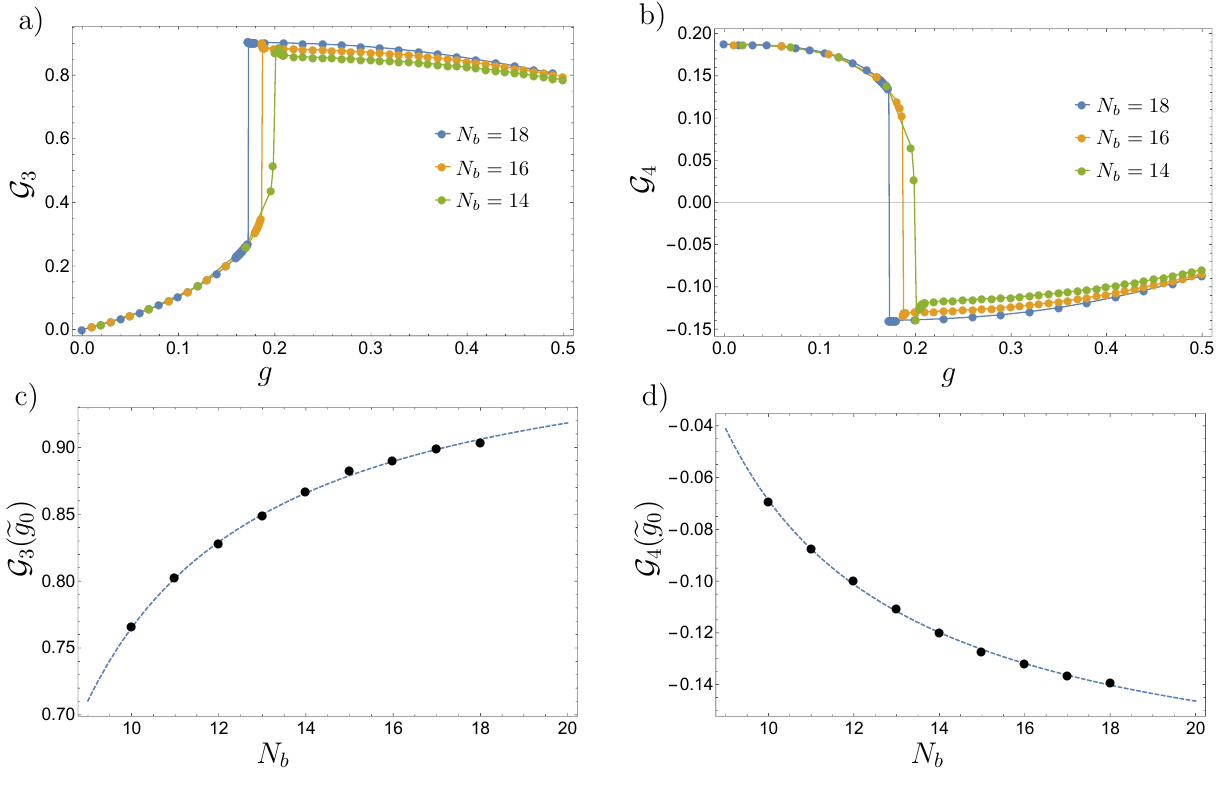}
\caption{a) $\mathcal{G}_3$ as a function of $g$ for different  $ N_{b}$.   b) $\mathcal{G}_4$ as a function of $g$ for different  $ N_{b}$.  c) The value of $\mathcal{G}_3 (g_0)$ vs $N_{b}$. d) The value of $\mathcal{G}_4 (g_0)$ vs $N_{b}$.  The black dots in panel c and d represent the numerical data using Rayleigh-Ritz method and the blue dashed lines are fits in Eq.(\ref{g3g4_fit}). Here, we have chosen $c=0$. }\label{Fig_8} 
\end{center}
\end{figure}

There is simple intuitive explanation for the existence of the critical point $g_0^\ast$ and this sudden change in $\Phi$.
For very small $g$,  perturbation theory around $g=0$ provides a reasonable estimate of the energy. To the leading order in $g$, we find
\begin{eqnarray}
E_{0}^{(2)}(g) = \frac{9}{2} -12 g^2 + O(g^4). \label{pert_gs_E}
\end{eqnarray}
Our numerical data  matches this perturbative estimate excellently $g<g^{\ast}_0$ (Fig.~\ref{Fig_1}a), but fails completely for  $g>g^{\ast}_0$. The location of the discontinuity tentatively identifies the end of the perturbative regime. 

In this regime, as the value of $\Phi$ is very small, we can conclude that the particle is effectively trapped in the well around $A_i=0$. This is a consequence of the high potential barrier between the two wells, and suppression of tunnelling  as discussed in Sec.~\ref{sec_2}.  The probability of finding the particle at the saddle point (or for that matter, any configuration with equal singular values) is zero.  As a result, the ground state $|\Psi_{0}^{(2)}\rangle$ remains a normalizable wavefunction in $\mathcal{H}_{phys}$ for $g<g_{0}^{\ast}$.

A simple estimate of the validity of the perturbative regime is provided by equating the ground state energy to the barrier height: 
\begin{eqnarray}
\frac{9}{2} -12g^2 =\frac{3}{32g^2},  \quad\quad \label{E0_eq_saddle}
\end{eqnarray}
giving us $ g \simeq 0.1488. $  This is remarkably close to the estimate of $g_0^\ast\simeq 0.143$ from FSS.

For all $g>g_{0}^{\ast}$, $\Phi$ is significantly higher. This implies that state $|\Psi_{0}^{(2)}\rangle$ has support  on configurations with large $A_i$ as well, and is no longer trapped in the well around $A_i=0$. This happens because the energy of $|\Psi_{0}^{(2)}\rangle$ for $g>g_{0}^{\ast} $ is higher than the barrier height,  and the particle is free to   travel back and forth between the potential wells.

Finally, let us consider $g=g_{0}^{\ast}$. The particle may travel between the two wells, but in order to do so must pass through the saddle point. This implies that the probability of finding the particle at the saddle point is non-zero. This  $|\Psi_{0}^{(2)}\rangle$ is not normalizable at $g=g_{0}^{\ast}$ and hence not a member of $\mathcal{H}_{phys}$.

This expulsion of  $|\Psi_{0}^{(2)}\rangle$ form $\mathcal{H}_{phys}$ is physically observed as a discontinuity in the energy at $g_{0}^{\ast}$. The true ground state of the $n_F=2$ sector at $g_{0}^{\ast}$ is now $|\Psi_{1}^{(2)}\rangle$, provided that it does not suffer from the same effect exactly at that same value of $g$. We will verify this when we will study the excited states in the following subsection.

At  $g_0^\ast$, the state $|\Psi^{(2)}_0\rangle$ is localized to the surface of equal singular values and is therefore conceptually like a translationally invariant state belonging to a non-regular representation. Further support for this claim comes from the expectation value of  the chromoelectric field squared $P_{ia}P_{ia}$, a manifestly positive operator. Finite size scaling  indicates that it scales to zero in the in the state $|\Psi^{(2)}_0\rangle$ at $g_0^\ast$ (Fig.~\ref{Fig_electric field}). This is  a ``condensate'' and the vanishing of electric field naturally suggests its identification as the dual superconducting phase.

 \begin{table}
\begin{center}
\begin{tabular}{|c||c|c|c|c|c|c|c|c|c}
\hline 
Observable & $C_{A,0}$ & $\alpha_{A,0}$ & $\beta_{A,0}$   & $\gamma_{A,0}$ \\ \hline  \hline  
$\mathcal{O}_{1,0}$ & -14.77 & 1.19 & 1.39 & 2.2  \\ 
& & & & \\ 
$\mathcal{O}_{2,0}$ &3.45 &   0.06  & 2.45& 2.1 \\ 
& & & & \\ 
$\mathcal{O}_{3,0}$ & 0.36 &   0.08 &  1.86 & 2.09  \\ 
  \hline 
\end{tabular}
\caption{Parameters for the fits of $\mathcal{O}_{A,0}$ to the power-law dependance in (\ref{power_law_1}).}\label{Table-2}
\end{center}
\end{table}

Further evidence for the above explanation comes from Binder cummulants $\mathcal{G}_3$ and $\mathcal{G}_4$ 
 \begin{eqnarray}
{\mathcal{G}}_3 \equiv \frac{\sqrt{3}}{2} \frac{  \langle \Psi_{0}^{(2)}|\epsilon_{ijk} \epsilon_{abc} M_{ia} M_{jb} M_{kc}|\Psi_{0}^{(2)}\rangle  } { \langle \Psi_{0}^{(2)} | M_{ia} M_{ia}|\Psi_{0}^{(2)}\rangle ^{\frac{3}{2}}}, \quad\quad 
{\mathcal{G}}_4 \equiv \frac{9}{8} \left[   \frac{  \langle \Psi_{0}^{(2)} | M_{ib} M_{jc} M_{ic} M_{jb}|\Psi_{0}^{(2)}\rangle }{\langle \Psi_{0}^{(2)} | M_{ia} M_{ia}|\Psi_{0}^{(2)}\rangle ^{2}} -\frac{1}{2}  \right] \label{g_3_g_4_defn_2} 
\end{eqnarray}
which provide important insight about the localization properties of the state (Fig.~\ref{Fig_8}a and Fig.~\ref{Fig_8}b).  The values of $\mathcal{G}_3$ and $\mathcal{G}_4$ are constrained to lie within the ``arrowhead'' (see  Fig.~\ref{Fig_arrow_1} in appendix~\ref{app_arrowhead}). 
As $g \to 0$,  $({\mathcal{G}}_3,{\mathcal{G}}_4) \to (0, \frac{3}{16})$ which is the center of the arrowhead. In the perturbative regime with $g<g_{0}^{\ast}$, the values of $({\mathcal{G}}_3,{\mathcal{G}}_4)$ always lie in the bulk of the arrowhead.  As we approach $\widetilde{g}_0(N_{b})$, both  ${\mathcal{G}}_3$ and ${\mathcal{G}}_4$ jump discontinuously. Their limiting values can be  estimated using FSS as (Fig.~\ref{Fig_8}c-d)
\begin{eqnarray}
{\mathcal{G}}_3\Big|_{\widetilde{g}_0(N_{b})}\simeq 1  + \frac{1.24}{N_{b} -4.7 }, \quad\quad  {\mathcal{G}}_3\Big|_{\widetilde{g}_0 (N_{b})}\simeq -\frac{3}{16}  + \frac{0.63}{N_{b} -4.7 }.  \label{g3g4_fit}
\end{eqnarray}
Thus in the limit $N_{b} \to \infty$, as $\widetilde{g}_0 \to g_{0}^{\ast}$ we get $( {\mathcal{G}}_3, {\mathcal{G}}_4)\Big|_{g=g_{0}^{\ast}} \simeq (1, -\frac{3}{16})$,  i.e. the corner $B_+$ in Fig.\ref{Fig_arrow_1}.  This shows that as $ g \to g_{0}^{\ast}$, the state $|\Psi_{0}^{(2)}\rangle$ is dynamically driven to configurations with equal singular values.

\subsection{Excited states with $c=0$} 

\begin{figure}[t!]
\begin{center}
\includegraphics[width=16cm]{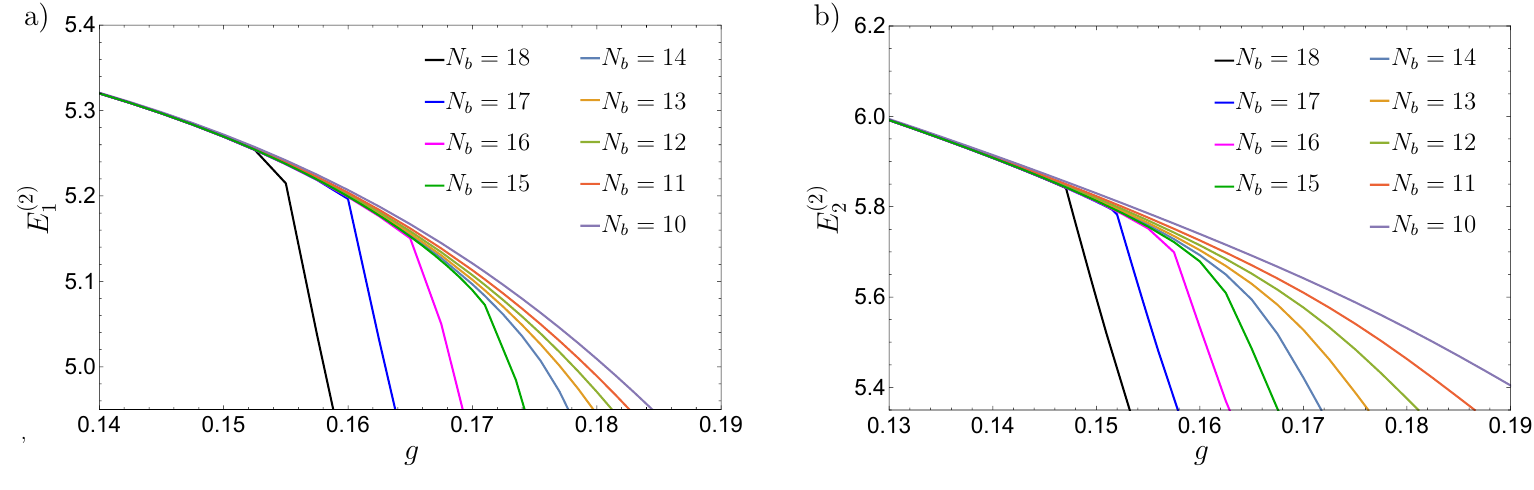}
\caption{ The kinks in the weak coupling regime for a) $E_1^{(2)}(g)$ and b) $E_2^{(2)}(g)$.  
Here, we have chosen $c=0$.}\label{Fig_excited_kink} 
\end{center}
\end{figure}

The discussion in the previous subsection focussed on the ground state of the $n_F=2$ sector. However, above mentioned effect  (i.e. the state being expelled from $\mathcal{H}_{phys}$ at a specific value of the coupling) is not an exclusive property of the ground state. The same phenomenon can be observed in the excited spin-0 states $|\Psi_{n}^{(2)}\rangle$  as well. For small $g$, the perturbative estimate of its energy is of the form
\begin{eqnarray}
E_{n}^{(2)} \approx a_n + b_n g^2 + O(g^4)
\end{eqnarray}
The coefficients $a_n$ and $b_n$ can be deduced by using perturbation theory or by fitting the numerical data near $g=0$.   Using these, we can estimate the value of $g_{n}^{\ast}$ at which $E_{n}^{(2)}(g)$ becomes equal to the barrier height at the saddle points: 
\begin{eqnarray}
a_n + b_ng_{n}^{\ast 2} = \frac{3}{32g_{n}^{\ast 2}}.  \label{En_eq_saddle}
\end{eqnarray}
These estimates of $a_n$, $b_n$ and $g_{n}^{\ast}$ for the first two excited states ($n=1,2$) are given in Table~\ref{Table-1}.

 \begin{table}
\begin{center}
\begin{tabular}{|c||c|c|c||c|c|c|c|c|c}
\hline 
Observable & $C_{A,1}$ & $\alpha_{A,1}$ & $\beta_{A,1}$   &  $C_{A,2}$ & $\alpha_{A,2}$ & $\beta_{A,2}$  \\ \hline  \hline  
$\mathcal{O}_{1,n}$ &-7.15&  0.250&  2.00 & -97.1& 7.77 & 1.10  \\    
& & & & && \\ 
$\mathcal{O}_{2,n}$ &4.09&   0.019& 2.87 & -9.47 &   0.018 & 2.97 \\  
& & & & && \\ 
$\mathcal{O}_{3,n}$ &1.68&   0.020& 2.37   & -4.37&   0.052& 2.16 \\   
  \hline 
\end{tabular}
\caption{Parameters for the power law fits of $\mathcal{O}_{A,n}$ to the power-law dependance in (\ref{power_law_2}).}\label{Table-3}
\end{center}
\end{table}
The numerical simulations  confirm that   $E_{1}^{(2)}$ and $E_{2}^{(2)}$ also have kinks (Fig.~\ref{Fig_excited_kink}) and the data fails to converge near each of the kinks. The location of these kinks $ \widetilde{g}_n$ for different $N_b$ are shown in Fig.\ref{Fig_gnstar}, and  FSS shows that  
\begin{eqnarray}
\widetilde{g}_n(N_{b}) \simeq g_{n}^{\ast}   +  d_n\, e^{-k_n N_{b}}.  \label{fit_gn}
\end{eqnarray}
The parameters $g_{n}^{\ast}$, $d_n$ and $k_n$ are given in Table~\ref{Table-1}. The true location of these kinks are $g_{1}^{\ast} $ and $g_{2}^{\ast} $ whose values are in excellent agreement with the intuitive reasoning of (\ref{En_eq_saddle}). 
It should be noted that $g_{0}^{\ast}>g_{1}^{\ast}>g_{2}^{\ast}$.  

 \begin{figure}[t!]
\begin{center}
\includegraphics[width=16cm]{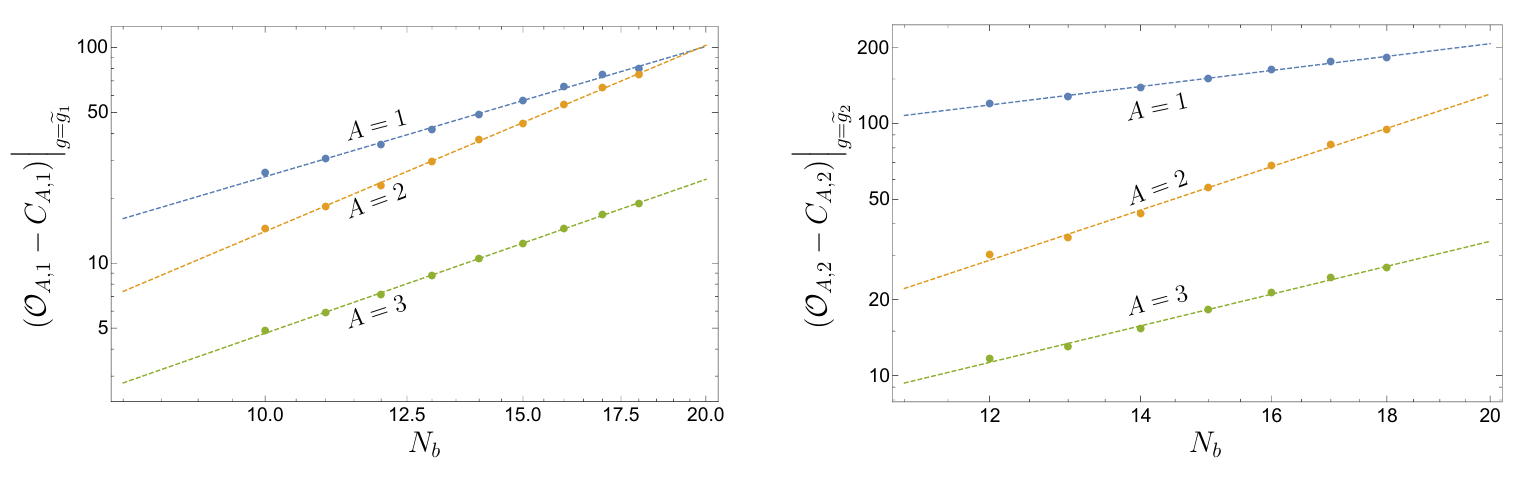}
\caption{Log-log plot of $(\mathcal{O}_{A,1}-C_{A,1})\Big|_{g=\widetilde{g}_1}$ and $(\mathcal{O}_{A,2}-C_{A,2})\Big|_{g=\widetilde{g}_1}$ vs $N_{b} $.  The dots  represent the numerical data using Rayleigh-Ritz method and the dashed lines are  fits
in (\ref{power_law_2}) with parameters in Table~\ref{Table-3}.  Here, we have chosen $c=0$.  }\label{Fig_11_new} 
\end{center}
\end{figure}

Further, we find that the  expectation values  $\mathcal{O}_{A,1}$ and $\mathcal{O}_{A,2}$ (in the excited states at $\widetilde{g}_1$ and $\widetilde{g}_2$) also have power-law dependence (Fig.~\ref{Fig_11_new}):  
\begin{eqnarray}
 \mathcal{O}_{A,n}\Big|_{g=\widetilde{g}_n}= C_{A,n}+  \alpha_{A,n} (N_{b})^{\beta_{A.n}}, \quad\quad n=1,2. \label{power_law_2}
\end{eqnarray}
The corresponding parameters are given in Table~\ref{Table-3}.  These diverge as $N_b \to \infty$ at their respective critical points.

 This argument suggests that in general, the state $|\Psi_n^{(2)}\rangle$ does not vanish at the saddle point at $g_{n}^{\ast}$ and hence is non-normalizable. There are thus an infinite number of isolated singular points $g_{0}^{\ast}<g_{1}^{\ast}<g_{2}^{\ast}< \ldots$ accumulating  at $g=0$.  At each of these singular points, one state in the $n_F=2$ sector is expelled from the $\mathcal{H}_{phys}$. As a consequence, each $E_n^{(2)}$ (and expectation values $\mathcal{O}_{A,n}$) has a discontinuity (and divergences) for $g\leq g_{0}^{\ast}$.

Following the arguments in the previous subsection, we can thus identify the state $|\Psi_n^{(2)}\rangle$ at $g_{n}^{\ast}$ as a dual superconducting phase as well.

\subsection{Non-zero $c$} 
\begin{figure}[b!]
\begin{center}
\includegraphics[width=8cm]{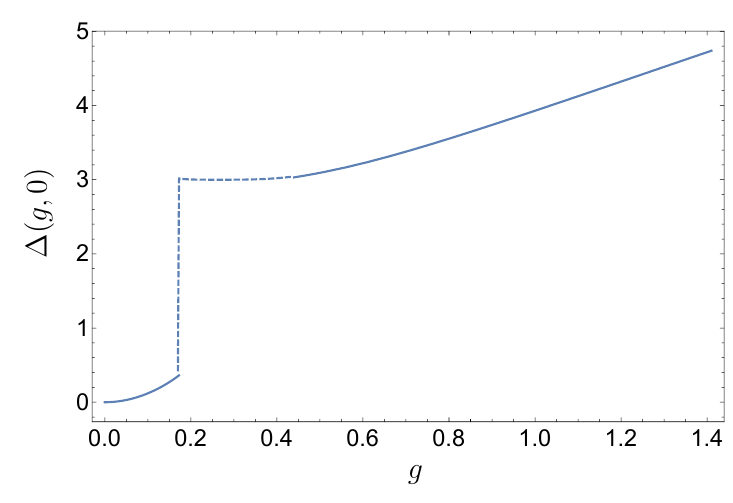}
\caption{The energy gap  $\Delta(g,0)$ as a function of $g$. Here, we have used the data from Rayleigh-Ritz method with extrapolation for $N_{b} \to \infty$ limit as in Fig.~\ref{Fig_1}.   }\label{Fig_energy_gap} 
\end{center}
\end{figure}

When $c\neq 0$, the energy of an eigenstate $|\Psi_n^{(n_F)}\rangle$ of Hamiltonian (\ref{Ham_1})  becomes
\begin{eqnarray}
E_n^{( n_F)}(g,c) = E_n^{ (n_F)}(g, 0) +  c(n_F-3).  
\end{eqnarray}
We now need to compare the energies ${E}_0^{(0)}(g,c)$  with $E_0^{(2)}(g,c)$ to decide whether  $|\Psi_0^{(2)}(g)\rangle$ or $|\Psi_0^{(0)}(g)\rangle$ has lower energy.

Let  $\Delta (g,c)$ be the gap 
\begin{eqnarray}
\Delta (g,c) \equiv {E}_0^{(0)}(g,c)- E_0^{(2)}(g,c) = \Delta(g,0)-2c.
\end{eqnarray}
$ \Delta(g,0)$ is shown in Fig.~\ref{Fig_energy_gap}.

For a given $g$,  there exists a $c$ such that $\Delta(g,c)=0$. We denote that value of $c$ as $c^R$: 
\begin{eqnarray}
c^R(g) =  \frac{1}{2} \Delta(g,0). 
\end{eqnarray}
The ground state is two-fermion state $|\Psi_0^{(2)}\rangle$ for $c<c^R(g)$ and it changes to pure glue state $|{\Psi}_0^{(0)} \rangle$ for $c>c^R(g)$.

Conversely, for any fixed $c>0$, there exists a value of the coupling $g_0^R(c)$ such that  $\Delta(g_0^R,c)=0$. The ground state changes across $g=g_0^R$ showing that there are two phases:
\begin{eqnarray}
|\Psi_{gs}(g,c)\rangle&=&|\Psi_0^{0,0} (g)\rangle, \quad\quad \text{for } g<g_0^R(c),  \quad\quad \text{phase I} \nonumber  \\
&=& |\Psi_0^{0,2}(g)\rangle,  \quad\quad \text{for } g>g_0^R(c),  \quad\quad \text{phase II}.\label{phase_eq_1}
\end{eqnarray}
At $g_0^R(c)$, there is a level crossing in the ground state and the system undergoes a QPT.  It is easy to see that the first derivative of the ground state energy $E_{gs}(g,c)$ is discontinuous at $g_0^R$: 
\begin{eqnarray}
\frac{\partial E_{gs}}{\partial c}  &=&-3, \quad\quad \text{for } g<g_0^R(c),  \quad\quad \text{phase I}  \nonumber  \\
&=& -1,  \quad\quad \text{for } g>g_0^R(c), \quad\quad \text{phase II}  \label{phase_eq_2}
\end{eqnarray}
Thus the level crossing at $g_0^R(c)$ is a first order phase transition.  
\begin{figure}[t!]
\begin{center}
\includegraphics[width=18cm]{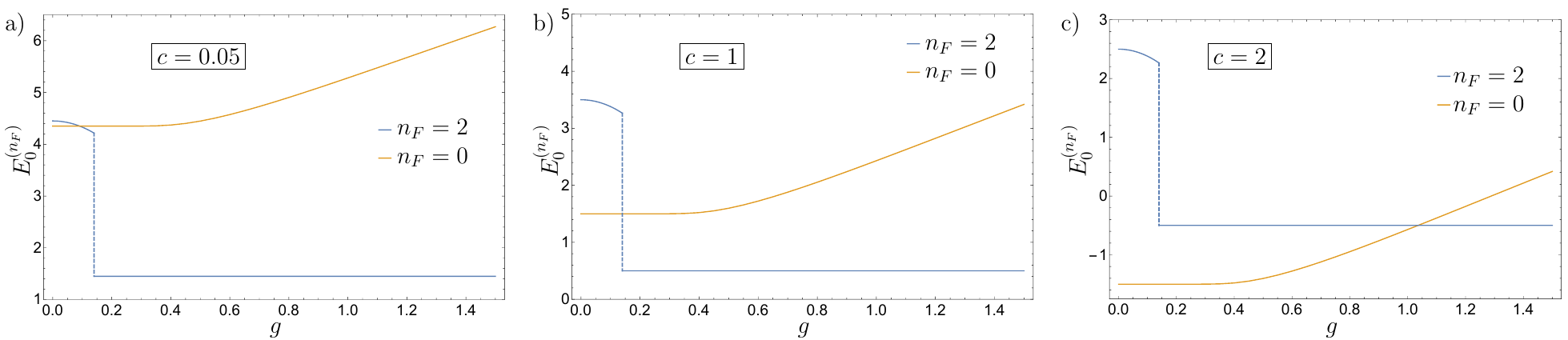}
\caption{$E_0^{(2)}(g,c)$ and ${E}_0^{(0)}(g,c)$ as function of $g$ for three different values of $c$. a) $c=0.05$ is an example of case I. b) $c=1$ is an example of case II. c) $c=2$ is an example of case III. Here, we have used the data from Rayleigh-Ritz method with extrapolation for $N_{b} \to \infty$ limit as in Fig.~\ref{Fig_1}.   }\label{Fig_13} 
\end{center}
\end{figure}
\begin{figure}[h]
\begin{center}
\includegraphics[width=18cm]{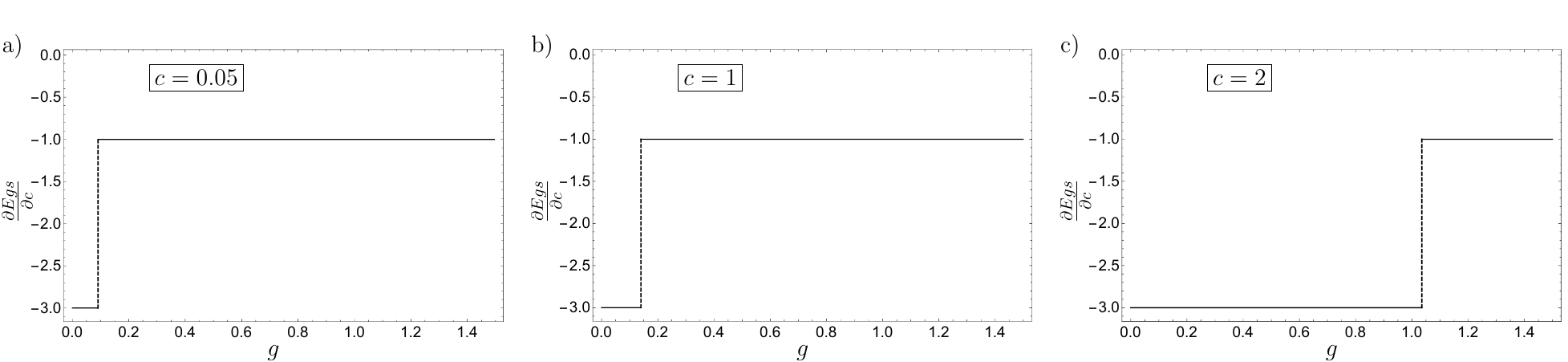}
\caption{$(\partial E_{gs}/\partial c)$ as function of $g$ for three different values of $c$. a) $c=0.05$ is an example of case I. b) $c=1$ is an example of case II. c) $c=2$ is an example of case III.    }\label{Fig_10} 
\end{center}
\end{figure}

The energy of the state $|\Psi_0^{(0)}\rangle$ for small $g$  can be obtained from (\ref{fit_E0_nf0}): $E_0^{(0)} \simeq 9/2 - 3c + O(\text{exp}(-1/g^2))$. For $g>g_{0}^{\ast}$, the energy of $|\Psi_0^{(2)}\rangle$ is $E_0^{(2)} \simeq 3/2 - c$. Thus for a level crossing when $g_0^R>g_{0}^{\ast}$ it is necessary  $9/2 - 3c<3/2 -c$. 
On the other hand, for $g<g_{0}^{\ast}$, the energy of $|\Psi_0^{(2)}\rangle$ is $E_0^{(2)} \simeq 9/2 -12 g^2- c$. Hence for a level crossing to occur when $g_0^R<g_0^{\ast}$, we require that $9/2 - 3c>9/2-12 g_{0}^{\ast^2} -c$.

\begin{figure}[b]
\begin{center}
\includegraphics[width=7cm]{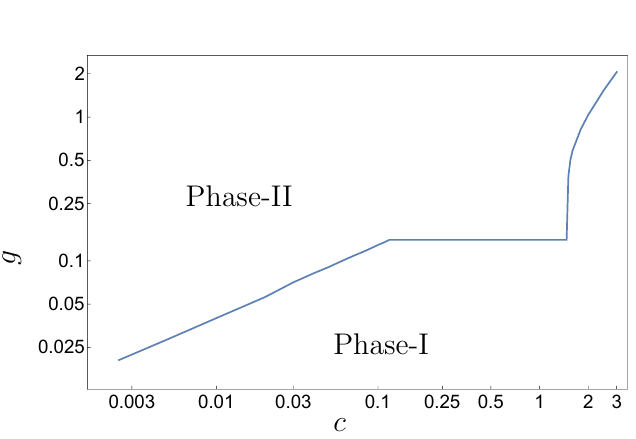}
\caption{Phase diagram in $c-g$ plane.  The solid blue line denotes $g_0^R(c)$ where both phases can co-exist.   }\label{Fig_11} 
\end{center}
\end{figure}

Depending on $c$, there are three possibilities for the location of QPT: 
\begin{enumerate}
\item case I -- when $g_0^R(c)< g_{0}^{\ast}$. This requires  $0\leq c<6 g_{0}^{\ast^2}\approx 0.132$. An example of this case is shown in  Fig.~\ref{Fig_13}a and the corresponding $\partial E_{gs}/\partial c$ in shown in Fig.~\ref{Fig_10}a. 
\item case II -- when $g_0^R(c) =g_{0}^{\ast}$. Now  $6 g_{0}^{\ast^2}\leq c<\frac{3}{2}$. An example of this case is shown in  Fig.~\ref{Fig_13}b and the corresponding $\partial E_{gs}/\partial c$ in shown in Fig.~\ref{Fig_10}b. 
\item case III -- when $g_0^R(c) >g_{0}^{\ast}$. This happens when $c>\frac{3}{2}$.  An example of this case is shown in  Fig.~\ref{Fig_13}c and the corresponding $\partial E_{gs}/\partial c$ in shown in Fig.~\ref{Fig_10}c. 
\end{enumerate}
 Fig.~\ref{Fig_11} shows  the phase diagram in the $c-g$ plane,  where the solid blue line denotes $g_0^R(c)$ where both phases can co-exist.

Interestingly, it is only in cases I and II that the ground state energy has a discontinuity at $g_{0}^{\ast}$: it gets localized near $A_i=0$ for $g<g_0^\ast$. In contrast for case III, the ground state energy is always continuous (in $g$) and the wavefunction never gets localized on the surface of equal singular values. Thus in case III, the ground state never becomes  a dual superconductor. 

The energies of the excited states (with $n_F=2$) have discontinuities at $g_{n}^{\ast}$ in all the cases.

\subsection{ SUSY breaking for $g<g_0^\ast$}

When  $c=1$, the  theory formally has $\mathcal{N}=1$ supersymmetry generated by $\mathcal{Q}_{\alpha}$ \cite{ErrastiDiez:2020iyk} (see Appendix \ref{app_sym} for explicit forms of the super charges).  This belongs to the case II discussed above, where the level crossing QPT coincides with the isolated singular point of the ground state: $g_0^R(c=1)=g_{0}^{\ast}$. It is natural to ask what effect the isolated singularities $g_{n}^{\ast}$ have on supersymmetry. 

For any supersymmetric theory, the ground state $|\Psi_{gs}\rangle$ with strictly positive energy $E_{gs}$ must be a  SUSY-singlet.  For our case, $\mathcal{Q}_\alpha |\Psi_{gs} \rangle$ should vanish for all $\alpha$ and hence saturate the bound in (\ref{bound_1}), giving
\begin{eqnarray}
E_{gs} (g)  =   -\frac{1}{2} (n_F-3). \label{bound_2}
\end{eqnarray}
The supersymmetry is also manifest in the spectrum of excited states as degenerate supermultiplets, each containing two bosonic (integer spin) states and a fermionic (half-integer spin) doublet. 
\begin{figure}
\begin{center}
\includegraphics[width=7cm]{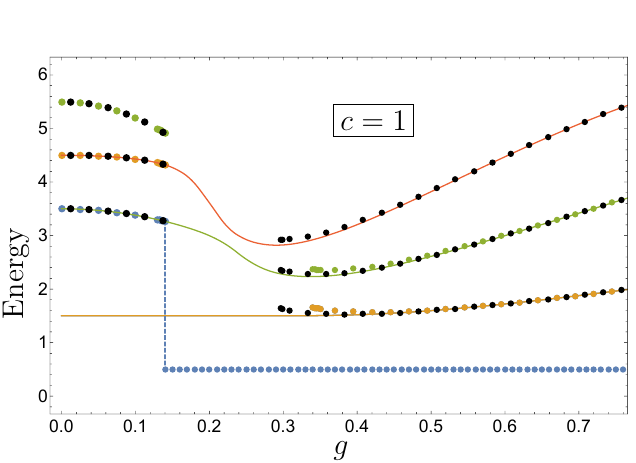}
\caption{ Energies of a few low-lying  spin-0 (colored line and dots) and  spin-1/2 (black dots) as functions of $g$ for $c=1$.  The unpaired state with the lowest energy is the ground state.  The degenerate excited states form  multiplets. Each spin-1/2 state is a doublet. 
}\label{Fig_12} 
\end{center}
\end{figure}

As we observed in the previous section, the ground state for $c=1$ is always unique. For $g<g_{0}^{\ast}$, the ground state energy is $E_{gs} \simeq 9/2-3c =3/2$  and $n_F=0$ (there are order $O(\text{exp}(-1/g^2))$ corrections to $E_{gs}$ which are negligible for small $g$ and do not affect the physics). On the other hand, for $g>g_{0}^{\ast}$, the ground state energy is $E_{gs} \simeq 3/2-c =1/2$ and $n_F=2$. Thus (\ref{bound_2}) is satisfied for both $g<g_{0}^{\ast}$ and $g>g_{0}^{\ast}$. Hence the ground state is a unique in both phases.

Now let us examine the supermultiplets. The first few multiplets are shown in Fig.~\ref{Fig_12}.  A typical putative SUSY multiplet consists of spin-0 states  $(|\Psi_{n}^{(0)}\rangle, |\Psi_{m}^{(2)}\rangle)$, and a spin-1/2 doublet  $|\Psi_{k, 1/2}^{(1)}\rangle$, where $m$, $n$ and $k$ label the energy levels. 
If the degeneracy between the bosonic states is lifted, then SUSY is broken. This is precisely what happens to these multiplets at each $g_{n}^{\ast}$: 
 the  state $|\Psi_n^{(2)}\rangle$ is expelled from $\mathcal{H}_{phys}$. As a result, one multiplet  has a missing bosonic state at $g_{n}^{\ast}$.  Thus SUSY is broken at each of the singular points $g_{n}^{\ast}$.

What is the status of SUSY  in the intervals $ g_{n+1}^{ \ast} <g<g_{n}^{\ast}$?
For $g$ infinitesimally smaller than $g_{n+1}^{ \ast}$,  consider the bosonic sector  $( |\Psi_{n+1}^{(2)}\rangle,  |\Psi_{n'}^{(0)}\rangle )$  of the multiplet. At $ g_{n+1}^{ \ast}$, the energy of the state $|\Psi_{n+1}^{(2)}\rangle$ is discontinuous, while the energy of $|\Psi_{n'}^{(0)}\rangle$ remains continuous. As a consequence of this discontinuity,  the energy of  $|\Psi_{n+1}^{(2)}\rangle$  is lower than that of $|\Psi_{n'}^{(0)}\rangle$ for $g>g_{n+1}^{ \ast}$. This is clearly visible in Fig.~\ref{Fig_12} as the pairing of the partner states changes in the vicinity of the critical point.  Thus the state $|\Psi_{n+1}^{(2)}\rangle$ either becomes degenerate with the neighbouring  lower multiplet, or remains unpaired. If it is unpaired from its bosonic partner, of course SUSY is broken. If it becomes degenerate with this lower multiplet $( |\Psi_{n}^{(2)}\rangle,  |\Psi_{n''}^{(0)}\rangle )$, then that multiplet has three bosonic states in the range $ g_{n+1}^{ \ast} <g<g_{n}^{\ast}$.  Thus in either of the two situations, SUSY is broken between $ g_{n+1}^{\ast} <g<g_{n}^{\ast}$. This argument holds for every such interval. Consequently, the localized phase is non-supersymmetric while the delocalized phase has $\mathcal{N}=1$ SUSY.

\section{Discussion}
We have shown that  the  weak coupling regime of matrix-QCD$_{2,1}^{\text{adj}}$ in the chiral limit exhibits a rich phase structure. We find that there is a localization-delocalization transition at $g_0^\ast$ with profound implications for the symmetry properties of the ground state. 
The non-normalizable state at $g_0^\ast$ corresponds to a non-regular representation of the Heisenberg-Weyl algebra and points to the formation of a condensate. The numerical evidence for this is the vanishing of $\text{Tr}(\Pi_i \Pi_i)$, as well as the divergence of $\text{Tr}(A_i A_i)$ in this state. This also suggests a nice physical interpretation that this state corresponds to a dual superconductor.

For non-zero $c$, there is  an additional first order QPT, where the fermion content of the ground state changes.   For $c>3/2$, there is no localization-delocalization transition in the ground state but the first order transition survives. However,  the localization-delocalization phenomenon happens for any  excited state with $n_F=2$, irrespective of the value of $c$. In particular when $c=1$, this leads to spontaneous breaking of supersymmetry.

The non-normalizable states at $g_n^\ast$ correspond to the corner $B_+$ of the arrowhead (see appendix \ref{app_arrowhead}). Note that $B_+$ is actually the entire sub-manifold of equal singular values. 
Classically the dynamics at $g_0^\ast$ is  localized on the saddle point, i.e. the sphaleron. But in the quantum theory due to the nature of the non-regular representation, the wavefunction is totally spread out over $B_+$.  Thus an investigation of the quantum dynamics of these corner states becomes an interesting open question. 

The Hamiltonian (\ref{Ham_1}) restricted to $B_+$ is a candidate for understanding the dynamics of corner states. In this case, the glue spin and color are ``locked''  \cite{Pandey:2016hat}. Interestingly, the existence of the different phases for this restricted Hamiltonian was anticipated in \cite{ErrastiDiez:2020iyk, Pandey:2016hat}. There the tool of investigation was the improved Born-Oppenheimer approximation, which takes in to account an adiabatic connection and an adiabatic scalar potential induced in  the glue configuration space.

\mbox{}\\
\textbf{Acknowledgements: }  It is our pleasure to thank V. Parameswaran Nair, Dimitra Karabali and Abhishek Chowdhury for discussions.

\appendix

\section*{Appendices} 
\numberwithin{figure}{section}
\section{Symmetries of the system}\label{app_sym}
 In the total Hilbert space  $\mathcal{H}=\mathcal{H}_F \otimes \mathcal{H}_G$,  the spatial rotations on the glue and  the quark are generated by 
 \begin{eqnarray}
 &&  \mathcal{L}_i^{(1)} \equiv  -\epsilon_{ijk}  \Pi_{ja} M_{ka}, \quad  \mathcal{L}_i^{(2)} \equiv  \frac{1}{2} b^\dagger_{\alpha a} \sigma^i_{\alpha \beta } b_{\beta a}, \\
&&[ \mathcal{L}_i^{(\alpha)} ,  \mathcal{L}_j^{(\beta)} ] = i  \delta_{\alpha\beta} \epsilon_{ijk} \mathcal{L}_k^{(\alpha)}, \quad  i,j,k=1,2,3,\quad  \alpha, \beta =1,2.
 \end{eqnarray}
The gauge rotations  on the glue and the quark  are generated by 
 \begin{eqnarray}
 && G_a^{(1)}\equiv -\epsilon_{abc} \Pi_{ib} M_{ic}, \quad G_a^{(2)}\equiv -i \epsilon_{abc} b^\dagger_{\alpha b}  b_{\alpha c}, \\
 && [ G_a^{(\alpha)} , G_b^{(\beta)} ] = i  \delta_{\alpha\beta} \epsilon_{abc}G_c^{(\alpha)}, \quad  a,b,c=1,2,3,\quad  \alpha, \beta =1,2. \label{gauss_law1}
 \end{eqnarray}

 The Hamiltonian (\ref{Ham_1}) does not commute with $ \mathcal{L}_i^{(\alpha)}$ or $G_a^{(\alpha)}$. It only commutes with total spin $J_i \equiv  \mathcal{L}_i^{(1)} +  \mathcal{L}_i^{(2)} $ and $G_a \equiv G_a^{(1)} + G_a^{(1)} $: 
  \begin{eqnarray}
 [J_i, J_j]=i\epsilon_{ijk} J_k, \quad\quad [G_a, G_b]=i \epsilon_{abc} G_c, \quad\quad [H, J_i]=0=[H, G_a]. \label{gauss_law2}
 \end{eqnarray}
 The  $J_i$'s generate the spatial rotation group $SO(3)_{rot}$ while $G_a$'s are the generators of the Gauss' law constraint. 
 
  The Gauss law constraint demands that all physical observables  commute with $G_a$ and that the physical Hilbert space $\mathcal{H}_{phys}$ is the color-singlet subspace of $\mathcal{H}$. Thus any $| \Psi \rangle \in \mathcal{H}_{phys}$ satisfies  $G_a | \Psi \rangle =0$ \cite{Pandey:2019dbp}. 

Under gauge transformations $h \in SU(2)$, the gauge field transforms as $A_i \to A_i' = h^{-1} A_i h$. The glue configuration space $\mathcal{C}_G$ is the set of all  gauge inequivalent  $M_{ia} \in M_3(\mathbb{R})$. As a result, $\mathcal{C}_G =M_3(\mathbb{R})/ Ad(SU(2))$ is twisted \cite{Singer:1978dk, Narasimhan:1979kf} -- the relic  in the matrix model of the Gribov problem.

\textit{Global $U(1)_R$ symmetry:}  In matrix-QCD$_{2,1}^{adj}$, the $U(1)_R$ transformations are generated by $Q_R$ which classically commute with the Hamiltonian $H$ in (\ref{Ham_1}). However in the quantum theory, the axial anomaly breaks the classical $U(1)_R$ symmetry \cite{Acharyya:2021egi}. Nonetheless, $Q_R$ remains a well-defined operator in $\mathcal{H}_{phys}$ and  the term $c \,Q_R$ in the Hamiltonian 
can be called, with some qualifications \cite{Braguta:2015owi, Braguta:2016aov, Acharyya:2024pqj}, as the chiral chemical potential.

\textit{Supersymmetry:} It is well-known that if $c=1$, the matrix model has $\mathcal{N}=1$ supersymmetry \cite{Han:2019wue, ErrastiDiez:2020iyk}.  
We can define the fermionic operators $\mathcal{Q}_\alpha =  2b_{\beta a}^\dagger\sigma^i_{\beta \alpha} \text{Tr} \Big[(\Pi_{i} + i B_{i})T^a\Big]$  with $ \alpha =1,2$ and a bosonic operator $Q_R=b^\dagger_{\alpha a} b_{\alpha a}-3=n_F-3$. It is straightforward to show that 
\begin{eqnarray}
[H, \mathcal{Q}_\alpha ] = i g b_{\alpha a}^\dagger G_a + (c-1) 
\mathcal{Q}_\alpha, \quad\quad [H, Q_R]=0, \quad\quad [Q_R, \mathcal{Q}_\alpha]=\mathcal{Q}_\alpha. \label{SUSY_rel_1}
\end{eqnarray}
As is evident from (\ref{SUSY_rel_1}),  the  Hamiltonian (\ref{Ham_1}) commutes with $\mathcal{Q}_\alpha$ in $\mathcal{H}_{phys}$ when $c=1$. Thus, the chiral matrix-QCD$_{2,1}^{adj}$ with $c=1$ has $\mathcal{N}=1$ supersymmetry generated by the supercharges $\mathcal{Q}_\alpha$.  

\textit{Hamiltonian from $\mathcal{Q}_\alpha$  and $\mathcal{Q}_\alpha^\dagger$:} The fermionic operators $\mathcal{Q}_\alpha$ have an implicit dependence on the Yang-Mills coupling  $g$ through the $B_i$.  For any given value of $g$, be  it the supersymmetric case (i.e. $c=1$) or not (i.e. $c \neq 1$), we can write the  Hamiltonian (\ref{Ham_1}) using  $\mathcal{Q}_\alpha$  and $\mathcal{Q}_\alpha^\dagger$.  For any given value of $g$,   the operators $\mathcal{Q}_\alpha(g)$  and $\mathcal{Q}_\alpha^\dagger(g)$ obey
\begin{eqnarray}
 \{ \mathcal{Q}_\alpha(g),  \mathcal{Q}_\beta^\dagger(g) \} = 2\delta_{\alpha \beta} \Big[  \rho H(g) +     \Big(\frac{3}{2}-c\Big) Q_R  \Big] {- 2\sigma^i_{\beta\alpha }  (J_i + M_{ia} G_a)}. 
\end{eqnarray}
Thus the Hamiltonian of matrix-QCD$_{2,1}^{adj}$ for any $c$ and $g$ can be expressed as 
\begin{eqnarray}
H(g)= \frac{1}{4 \rho} \{ \mathcal{Q}_\alpha(g),  \mathcal{Q}_\alpha^\dagger(g) \}  +  \frac{1}{\rho}  \Big(c-\frac{3}{2}\Big) Q_R. 
\end{eqnarray}
As  $ \{ \mathcal{Q}_\alpha,  \mathcal{Q}_\alpha^\dagger \} $ is semi-positive definite,  for any $|\Psi \rangle \in \mathcal{H}_{phys}$,  the expectation value $\langle \Psi| \{ \mathcal{Q}_\alpha,  \mathcal{Q}_\alpha^\dagger \} |\Psi\rangle \geq 0$. This leads to the inequality $ \langle \Psi| H |\Psi \rangle   \geq   (2\rho)^{-1} (2c-3) \langle \Psi| Q_R |\Psi \rangle $.  
Further, if $|\Psi_n \rangle \in  \mathcal{H}_{phys}$ is an eigenstate of $H$ satisfying $H|\Psi_n \rangle= \rho^{-1} E_n |\Psi_n \rangle$, the above leads to an lower bound on the energy of the ground state $|\Psi_0 \rangle$: 
\begin{equation}
E_0 (g)  \geq    \Big(c-\frac{3}{2}\Big) \langle \Psi_n | Q_R |\Psi_n \rangle \label{bound_1}
\end{equation}
This has been explicitly verified in Fig.~\ref{Fig_En_nf0_sector} and Fig.~\ref{Fig_1}.

\section{Minima of $V_{YM}$ in Rectangular Coordinates}\label{app_min_rect}
The Yang-Mills potential has two minima at $M_{ia}^{min}=0, \delta_{ia}/g$ which are separated by a barrier at the saddle point $M_{ia}^{saddle}= \delta_{ia}/(2g)$. The $V_{min}=0$ for both minima and the saddle height is $3/(32 g^2 \rho)$. 
For very small $g$, we find that the ground state and few low-energy excited states are localized in the well around $M_{ia}^{min}=0$. This may seem slightly puzzling: why does the ground state prefer to stay localized near $M_{ia}=0$, though $V_{min}=0$ for both wells?

For extremely small $g$, when the tunnelling is suppressed, we can understand the nature of the ground state as follows.  Let us  expand the potential around $M_{ia}^{min}$ by defining $M_{ia} = M_{ia}^{min}+ m_{ia}$.  For small $g$, the leading order effective potential is given by 
\begin{eqnarray}
V_{eff} \approx \frac{1}{2} \frac{\partial^2V_{YM}}{\partial M_{ia} \partial M_{jb} }\Big|_{M_{ia}^{min}}  m_{ia} m_{jb} + \text{higher order terms}. 
\end{eqnarray} 
where the first term is independent of $g$, and the higher order terms are $O(g)$ or higher.  Due to this quadratic term, the zero-point energy is the leading contribution to the  energy of any eigenstate of $H$.

First, let us  consider what happens at $g=0+\epsilon$.
It is straightforward to see that around $M_{ia}^{min}=0$, we have 
\begin{eqnarray}
V_{eff} &\approx& \frac{1}{2} \frac{\partial^2V_{YM}}{\partial M_{ia} \partial M_{jb} }\Big|_{m_{ia}=0} m_{ia} m_{jb} \approx \frac{1}{2} m_{ia} m_{ia}
\end{eqnarray}
and the excitations of this oscillator system have energies $(n+9/2)$ with $n=0,1,2\cdots$.

On the other hand, for  $M_{ia}^{min}= \delta_{ia}/ \epsilon$ the effective potential is
\begin{eqnarray}
V_{eff} &\approx& \frac{1}{2}\Big[ \frac{1}{3}(m_{11}+m_{22}+m_{33})^2 + 2(m_{11}-m_{22})^2+ \frac{2}{3} (m_{11}+m_{22}-2m_{33})^2+ \nonumber \\
&& \,\,\, 2(m_{12}+m_{21})^2 +  2(m_{13}+m_{31})^2+ 2(m_{23}+m_{32})^2\Big]+O({\epsilon})
\end{eqnarray}
Thus, the excitations of this oscillator system have energies $(n+11/2)$. However, as $m_{ia}=0$ lies on the surface of coincident singular values, the state with $n=0,2,\ldots$ are  not in $\mathcal{H}_{phys}$ as they do not vanish at $m_{ia}=0$. Therefore, the state which is localized near  $M_{ia}^{min}= \delta_{ia}/{\epsilon}$ has energy $ \geq 13/2$.

\section{Hamiltonian in the SVD coordinates}\label{app_SVD_1}

To study the quantum dynamics in the SVD coordinates, we start with the natural metric on the gauge configuration space
\begin{eqnarray}
ds^2 = \text{Tr} \Big[dM^T dM \Big].
\end{eqnarray}
Using the left-invariant one-forms on $SO(3)_{rot}$ and gauge group $AdSU(2) \sim SO(3)_{col}$: 
\begin{eqnarray}
R_0^T dR_0= -i \omega_{iR} T_i, \quad\quad S_0^T dS_0 = - i  \omega_{aS} T_a
\end{eqnarray}
we can write the metric in the SVD coordinates as
\begin{eqnarray}
ds^2= \sum_{i} \Big(da_i^2 + (X_2- a_i^2) (\omega_{iR}^2+  \omega_{iS}^2)-\frac{4 X_3}{a_i}  \omega_{iR} \omega_{iS}\Big), \quad X_2\equiv \sum_{k=1}^3 a_k^2, \quad X_3= a_1 a_2 a_3 \label{metric_bulk}
\end{eqnarray}
There are nine coordinates:  the three singular values $a_i$ and six angular variables $\omega_{iR}$ and $\omega_{iS}$. The measure for the inner product in the Hilbert space becomes \cite{Iwai:2010, Acharyya:2017uhl}
\begin{equation}
dV= \, d\Omega_R \, d\Omega_S \prod_i da_i  \label{measure_1}
\end{equation}
where $d\Omega_R$ and $ d\Omega_S$ are $SO(3)_{rot}$- and $SO(3)_{col}$-invariant volume forms.

The conjugate momenta for the $a_i$ coordinates are $(-\partial/\partial a_i)$ and for the angular coordinates are: 
\begin{eqnarray}
\widetilde{L}_{i} = (X_2- a_i a_i) \omega_{iR} - \frac{2 X_3}{a_i} \omega_{iS}, \quad\quad \widetilde{G}_{i} = (X_2- a_i a_i) \omega_{iS} - \frac{2 X_3}{a_i} \omega_{iR}. 
\end{eqnarray}

The kinetic term can be obtained from the  Laplace-Beltrami operator with the metric (\ref{metric_bulk}): 
\begin{eqnarray}
\hspace*{-0.5cm} \text{Tr} \Pi_i \Pi_i  &=&-\frac{1}{2} \Delta\\
&=&\sum_{i}\Big[ -\frac{1}{2}\frac{\partial^2}{\partial a_i^2}+ \frac{1}{4} \sum_{j\neq i}\frac{a_i^2 + a_j^2}{(a_i^2- a_j^2)^2}+   \frac{1}{4} \sum_{j\neq i, k \neq j, i} \frac{(a_j^2 + a_k^2) (\widetilde{L}_{i}^2+\widetilde{G}_{i}^2) + 4 a_j a_k \widetilde{L}_{i } \widetilde{G}_{i} }{(a_j^2 - a_k^2)^2} 
\Big]. 
\end{eqnarray}
To the above, if we add the potential and the interaction terms, we get the total Hamiltonian.

\subsection{Eigenstates of $H_{ref}$ vanishes on the surfaces $|a_i| =|a_j|$ }
We can construct the reference Hamiltonian
\begin{eqnarray}
H_{ref} = 
-\frac{1}{2} \Delta + \frac{1}{2} X_2
\end{eqnarray}
whose eigenstates gives a basis to expand the glue states in  $\mathcal{H}_{phys}$: 
\begin{eqnarray}
H_{ref} |\Psi\rangle = E |\Psi\rangle
\end{eqnarray}

We  can express $|\Psi\rangle$ as equivariant functions \cite{Iwai:2010}: 
\begin{eqnarray}
|\Psi\rangle = \sqrt{\phi} D^p(R_0) F \Big[D^q(S_0)\Big]^T e^{-\frac{1}{2} X_2}, \quad\quad  \phi \equiv (a_1^2-a_2^2)(a_2^2-a_3^2)(a_1^2-a_3^2)
\end{eqnarray}
where $D^p(R_0)$ and $D^q(S_0)$ are the Wigner $D$-matrices for $(2p+1)$- and $(2q+1)$-dimensional irreducible representations of $SO(3)_{rot}$ and $SO(3)_{col}$, respectively. Here, $F$ is a $(2p+1) \times (2q+1)$-dimensional matrix-valued  homogeneous function which is a zero mode of the Laplacian $\Delta$: 
\begin{eqnarray}
a_i \frac{\partial }{\partial a_i} F= n F, \quad\quad \Delta F=0, 
\end{eqnarray}
and satisfies the boundary condition 
\begin{eqnarray}
\lim_{a_j \to a_k} \Big[J_{i}^{(p)}J_{i}^{(p)} F+F  J_{i}^{(q)T}J_{i}^{(q)T} + 2 J_{i}^{p} F J_{i}^{(q)T} \Big]=0 \quad\quad i\neq j,k \label{bc_11}
\end{eqnarray}
where $J_{i}^{(p)}$  are the $(2p+1)$-dimensional spin matrices. 
It is easy to see that $|\Psi\rangle$ is square integrable wavefunction with measure (\ref{measure_1}) and gives finite energy. Also, because of the factor $\sqrt{\phi}$, the wavefunction $|\Psi\rangle$ vanishes on the surfaces $|a_i|=|a_j|$. 

Further, it is straightforward to see that 
\begin{eqnarray}
H_{ref} |\Psi\rangle &=& 
\sqrt{\phi}e^{-\frac{1}{2} X_2}D^p(R)\Big[ -\frac{1}{2}\Delta F \Big]\Big[D^q(S)\Big]^T  +
\Big(n+\frac{9}{2}\Big) |\Psi\rangle=  
\Big(n+\frac{9}{2}\Big) |\Psi\rangle. 
\end{eqnarray}
Therefore,  $\{|\Psi\rangle\}$ is the set of  energy eigenstates of the nine-dimensional harmonic oscillator which we have used to construct the basis for $\mathcal{H}_{phys}$.

\section{The Arrowhead}\label{app_arrowhead}

\begin{figure}
\begin{center}
\includegraphics[width=18cm]{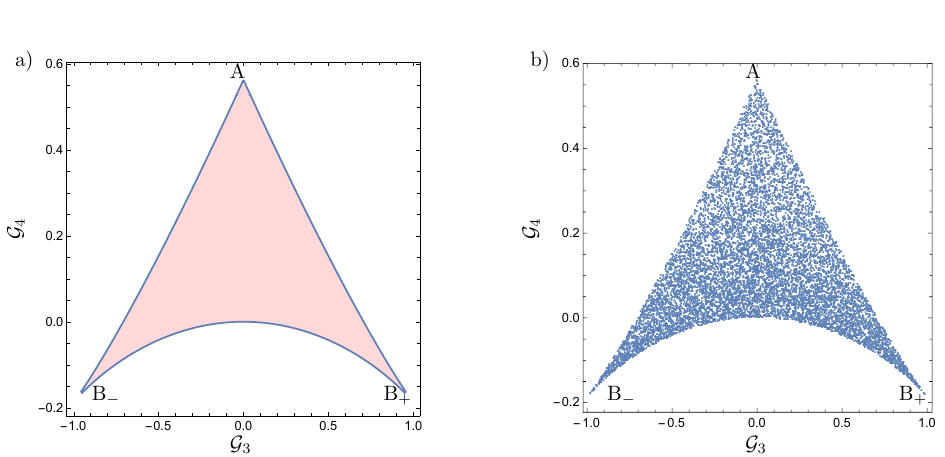} 
\caption{ a) In  $(\mathcal{G}_3 , \mathcal{G}_4)$ plane,  the shaded region is the classically allowed. b) $(\mathcal{G}_3 , \mathcal{G}_4)$ for 10000   real random matrix $M$ with  $-10^6 \leq M_{ia} \leq 10^6$.  }\label{Fig_arrow_1} 
\end{center}
\end{figure}

To distinguish the configurations on the surfaces $|a_i|=|a_j|>0$ from the others, it is useful to define the quantities
 \begin{eqnarray}
\mathcal{G}_3 \equiv \frac{\sqrt{3}}{2} \frac{  \epsilon_{ijk} \epsilon_{abc} M_{ia} M_{jb} M_{kc}  } { ( M_{ia} M_{ia} ) ^{\frac{3}{2}}}, \quad\quad 
\mathcal{G}_4 \equiv \frac{9}{8} \left[   \frac{  M_{ib} M_{jc} M_{ic} M_{jb} }{( M_{ia} M_{ia}) ^{2}} -\frac{1}{2}  \right] \label{g_3_g_4_defn_1} 
\end{eqnarray}
which in the SVD coordinates become 
\begin{eqnarray}
\mathcal{G}_3 \equiv \frac{ 3\sqrt{3}\,\,  a_1 a_2 a_3}{(a_1^2+ a_2^2 + a_3^2)^{\frac{3}{2}}}, \quad\quad \mathcal{G}_4 \equiv \frac{9}{8}  \left[   \frac{ a_1^4 + a_2^4 + a_3^4}{(a_1^2+ a_2^2 + a_3^2 ) ^{2}} -\frac{1}{2}  \right].
\end{eqnarray}
For any $3 \times 3$ real matrix $M_{ia}$, the  quantities $\mathcal{G}_3$ and $\mathcal{G}_4$ are constrained to lie inside the shaded ``arrowhead'' in Fig.\ref{Fig_arrow_1}a \cite{Pandey:2016hat}. To illustrate this, we have constructed 10000 real $M$ with randomly chosen $M_{ia} \in [-10^6, 10^6]$ and computed their  $\mathcal{G}_3$ and $\mathcal{G}_4$. As shown in Fig.\ref{Fig_arrow_1}a, the $\mathcal{G}_3-\mathcal{G}_4$ for these random matrices are constrained in the arrowhead \cite{Acharyya:2024pqj}.  

The boundary (the three edges and the three corners) of the arrowhead  is defined by:
 \begin{eqnarray}
&&  \text{at A}: \,\,|a_1| \neq 0, a_2=0=a_3, \quad\quad \quad\quad \text{at B}_\pm: \,\,a_1=a_2= \pm a_3, \nonumber \\
&&  \text{on AB}_\pm: \,\,a_1> a_2= \pm a_3, \quad\quad\quad\quad  \text{on B}_+\text{B}_-: \,\, a_1= a_2> a_3. 
 \end{eqnarray} 
 The interior points (bulk) of the arrowhead correspond to generic configurations (i.e. with distinct $|a_i|$), while as the boundaries correspond to those configurations with $|a_i|=|a_j|$. 
   In particular, the saddle point is at  $(\mathcal{G}_3, \mathcal{G}_4)=(1,-3/16)$, which corresponds to the corner $B_+$ in the arrowhead.

For the quantum  theory, we will the define the analogous quantities  in terms of expectation values of the operators. These are  essentially the third and the fourth Binder cumulants, whose analogues are extensively used in the study of spin systems.

\section{Non-regular representations of the Heisenberg-Weyl algebra} \label{nonregular_rep_HW_algebra}

Non-regular representations of the Weyl algebra were first discussed by \cite{Beaume:1974ad}. We provide here a quick review,  relying on the exposition by Acerbi et al \cite{Acerbi_combo}.

For an $n$-dimensional quantum mechanical system, the Heisenberg algebra is generated by 
$\widehat{x}_a$ and  $\widehat{p}_a$ ($a,b=1,2,\ldots n$) satisfying the canonical commutation relations (CCRs)   $[\widehat{x}_a, \widehat{p}_b]=i\delta_{ab}$. Alternately, one considers the unitary operators  $U(\alpha_a)$ and $V(\beta_a)$ ($\alpha_a$ and  $\beta_a$ are real parameters) satisfying the Weyl relation $U(\alpha_a)V(\beta_a)=e^{-i \alpha_a \beta_a} V(\beta_a)U(\alpha_a)$.

A fundamental question of quantum mechanics is to represent the Heisenberg (or Weyl) algebra irreducibly on some Hilbert space $\mathcal{H}$. If there is a representation wherein the operators $U$ and $V$ are weakly continuous in $\alpha_a$ and $\beta_a$, then the generators  $\widehat{x}_a$ and  $\widehat{p}_a$ are both self-adjoint operators defined  on  dense domains  $\mathcal{D}_x$ and $\mathcal{D}_p$, respectively.  In this case $\widehat{x}_a$ and  $\widehat{p}_a$ are both observables, implying  that  both  $\langle \psi| \widehat{x}_a^2 | \psi\rangle < \infty$ and  $\langle \chi| \widehat{p}_b^2 | \chi\rangle < \infty$ with $|\psi\rangle \in \mathcal{D}_x$ and $|\chi\rangle \in \mathcal{D}_p$. It is then straightforward to see that the Weyl relation is equivalent to the CCR. Such an irreducible representation indeed exists, and is called the regular representation. 

When the above conditions are satisfied, 
 a wavefunction  $\Psi(x_\alpha, t)$ in the Schr\"odinger representation satisfies 
\begin{eqnarray}
U(\alpha_a) \Psi(x_a, t)= e^{i \alpha_a x_a} \Psi(x_a, t), \quad\quad V(\beta_a) \Psi(x_a, t)=\Psi(x_a+ \beta_a, t). 
\end{eqnarray}
By the Stone-von Neumann theorem, any other unitary irreducible regular representation of the Weyl algebra is equivalent to the Schr\"odinger representation. Thus  the Schr\"odinger representation is the unique regular representation of the  Weyl algebra.

However, in certain physical situations, it may happen that although $U$ and $V$ exist as unitary operators, 
their generators may not exist as self-adjoint operators. A typical example of this is the physics of translationally invariant systems:  wavefunctions are no longer elements of $L^2(\mathbb{R}^n)$ and $\widehat{x}_a$ does not  exist as a self-adjoint operator \cite{Beaume:1974ad}. In this case the Weyl representation exists but not the Heisenberg one. This is a non-regular representation, not unitarily equivalent to the regular representation. 

There is a  simple  demonstration that the regular and non-regular representations are indeed inequivalent. Consider the partition function $Z_{box}$ of a particle in an $n$-dimensional box with each side $L$, and the Hamiltonian $\widehat{p}^2/2m$. When the box size $L$ is very large, the spacing between the energies is small enough for the sum to be approximated as an $n$-dimensional Gaussian integral. This gives $Z_{box}$, the average energy $\langle E \rangle_{box}$, and the specific heat capacity $C_{box}$ to be
\begin{eqnarray}
Z_{box}= \left(\frac{m L^2}{2 \pi \beta}\right)^{\frac{n}{2}}, \quad \langle E \rangle_{box}= - \frac{\partial \log Z_{box}}{\partial \beta}=\frac{n}{2} \beta^{-1}, \quad C_{box}= \frac{\partial  \langle E \rangle_{box}}{\partial T}= \frac{n}{2}
\end{eqnarray}
These are of course the well-known results. In particular for a free particle in contact with a thermal bath, the specific heat is independent of temperature.

\begin{figure}
\begin{center}
\includegraphics[width=18cm]{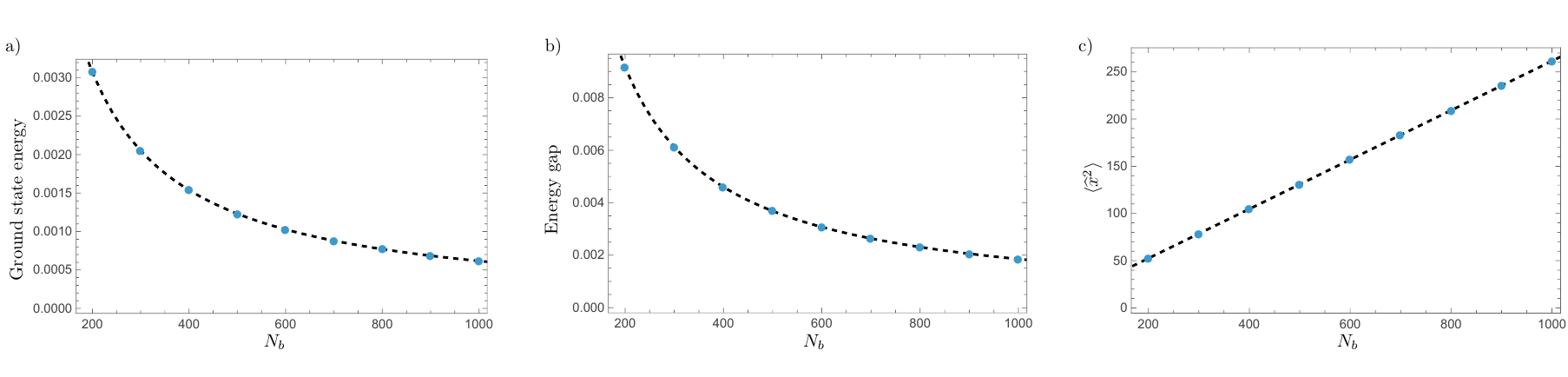} 
\caption{ One-dimensional translationally invariant system: a) the ground state energy, b) the energy gap, and c) $\langle \widehat{x}^2 \rangle$ as a function of the cut-off $N_b$ in the limit of $\beta \to \infty$ and $\omega_0 \to 0$. The blue dots represent the numerical data and the black dashed lines represent the fits: ground state energy $\sim 0.615 N_b^{-1}$, energy gap$\sim 1.83 N_b^{-1}$ and $\langle \widehat{x}^2 \rangle \sim 0.13 + 0.26 N_b$.  }\label{Fig_xsq_free_particle} 
\end{center}
\end{figure}

Alternately, we can consider an isotropic $n$-dimensional harmonic oscillator and compute the partition function, average energy, and specific heat:
\begin{eqnarray}
Z_{SHO}= \left(\frac{1}{2 \sinh \frac{\beta \omega_0}{2}}\right)^{n}, \quad \langle E \rangle_{SHO}= - \frac{\partial \log Z_{SHO}}{\partial \beta} =
\frac{n \omega_0}{2}\coth\frac{\beta \omega_0}{2}, \quad C_{SHO} = n \left( \frac{\beta \omega_0}{2 \sin \frac{\beta \omega_0}{2}}\right)^2
\end{eqnarray}
If we tune the oscillator frequency $\omega_0$ to zero, the Hamiltonian goes over to that of a free particle, but the  other thermodynamic quantities do not. 
In particular,
\begin{eqnarray}
\lim_{L \to \infty}C_{box} \neq  \lim_{\omega_0 \to 0}C_{SHO}.
\end{eqnarray}
proving the somewhat surprising result that $Z_{SHO}$ is not analytic  at $\omega_0=0$. This lack of analyticity arises because $Z_{SHO}$ is computed in the regular representation. In contrast,  the thermal average value  $\langle \mathcal{O} \rangle_{free\,\,particle} = \lim\limits_{L \rightarrow \infty} \langle  \mathcal{O}\rangle_{box}$ is computed in the non-regular translationally invariant representation. A mathematically rigorous discussion of this point is provided in \cite{Acerbi_combo}.

How does a non-regular representation show itself numerically? To this end, let us consider the translationally invariant Hamiltonian $H=\frac{\widehat{p}^2}{2m}$ in one dimension, and set out to "discover" its ground state energy and wavefunction by variational methods, using the first $N_b$ states of a 1-dimensional simple harmonic  oscillator. Using the ground state wavefunction, we can estimate $\langle\widehat{x}^2\rangle$.
The results are shown in Fig. \ref{Fig_xsq_free_particle}, where we have taken $N_b=1000$. The ground state energy and  the energy gap, as expected, vanishes as $1/N_b$. On the other hand,  $ \langle\widehat{x}^2\rangle \approx a+ b N_b$ with  $a\approx 0.13$ and $b\approx 0.26$. As a consequence, $ \langle\widehat{x}^2\rangle$ diverges in the $N_b \to \infty $ limit, showing that $\widehat{x}^2$ is ill-defined. This situation closely mimics what happens to the ground state wavefunction of our matrix model at the critical point $g_0^\ast$.


\begin{thebibliography}{99}



\bibitem{Balachandran:2014iya}
A.~P.~Balachandran, S.~Vaidya and A.~R.~de Queiroz,
Mod. Phys. Lett. A \textbf{30}, no.16, 1550080 (2015)
doi:10.1142/S0217732315500807
[arXiv:1412.7900 [hep-th]].

\bibitem{Balachandran:2014voa}
A.~P.~Balachandran, A.~de Queiroz and S.~Vaidya,
Int. J. Mod. Phys. A \textbf{30}, no.09, 1550064 (2015)
doi:10.1142/S0217751X15500645
[arXiv:1407.8352 [hep-th]].

\bibitem{Singer:1978dk}
I.~M.~Singer,
Commun. Math. Phys. \textbf{60}, 7-12 (1978)
doi:10.1007/BF01609471

\bibitem{Narasimhan:1979kf}
M.~S.~Narasimhan and T.~R.~Ramadas,
Commun. Math. Phys. \textbf{67}, 121-136 (1979)
doi:10.1007/BF01221361

\bibitem{Acharyya:2021egi}
N.~Acharyya, M.~Pandey and S.~Vaidya,
Phys. Rev. Lett. \textbf{127}, no.9, 092002 (2021)
doi:10.1103/PhysRevLett.127.092002
[arXiv:2104.04048 [hep-th]].

\bibitem{Acharyya:2016fcn}
N.~Acharyya, A.~P.~Balachandran, M.~Pandey, S.~Sanyal and S.~Vaidya,
Int. J. Mod. Phys. A \textbf{33}, no.13, 1850073 (2018)
doi:10.1142/S0217751X18500732
[arXiv:1606.08711 [hep-th]].

\bibitem{Pandey:2019dbp}
M.~Pandey and S.~Vaidya,
Phys. Rev. D \textbf{101}, no.11, 114020 (2020)
doi:10.1103/PhysRevD.101.114020
[arXiv:1912.03102 [hep-th]].


\bibitem{Pandey:2016hat}
M.~Pandey and S.~Vaidya,
J. Math. Phys. \textbf{58}, no.2, 022103 (2017)
doi:10.1063/1.4976503

\bibitem{Acharyya:2017uhl}
N.~Acharyya and A.~P.~Balachandran,
Phys. Rev. D \textbf{96}, no.7, 074024 (2017)
doi:10.1103/PhysRevD.96.074024
[arXiv:1702.06430 [hep-th]].


\bibitem{Acharyya:2024pqj}
N.~Acharyya, P.~Aich, A.~Bandyopadhyay and S.~Vaidya,
Phys. Rev. D \textbf{110}, no.5, 054016 (2024)
doi:10.1103/PhysRevD.110.054016
[arXiv:2406.06055 [hep-th]].





\bibitem{Witten:1981nf}
E.~Witten,
Nucl. Phys. B \textbf{188}, 513 (1981)
doi:10.1016/0550-3213(81)90006-7

\bibitem{Claudson:1984th}
M.~Claudson and M.~B.~Halpern,
Nucl. Phys. B \textbf{250}, 689-715 (1985)
doi:10.1016/0550-3213(85)90500-0


\bibitem{Cooper:1994eh}
F.~Cooper, A.~Khare and U.~Sukhatme,
Phys. Rept. \textbf{251}, 267-385 (1995)
doi:10.1016/0370-1573(94)00080-M
[arXiv:hep-th/9405029 [hep-th]].





\bibitem{Halpern:1997fv}
M.~B.~Halpern and C.~Schwartz,
Int. J. Mod. Phys. A \textbf{13}, 4367-4408 (1998)
doi:10.1142/S0217751X98002110
[arXiv:hep-th/9712133 [hep-th]].


\bibitem{Danielsson:1996uw}
U.~H.~Danielsson, G.~Ferretti and B.~Sundborg,
Int. J. Mod. Phys. A \textbf{11}, 5463-5478 (1996)
doi:10.1142/S0217751X96002492
[arXiv:hep-th/9603081 [hep-th]].




\bibitem{Ishibashi:1996xs}
N.~Ishibashi, H.~Kawai, Y.~Kitazawa and A.~Tsuchiya,
Nucl. Phys. B \textbf{498}, 467-491 (1997)
doi:10.1016/S0550-3213(97)00290-3
[arXiv:hep-th/9612115 [hep-th]].

\bibitem{Banks:1996vh}
T.~Banks, W.~Fischler, S.~H.~Shenker and L.~Susskind,
Phys. Rev. D \textbf{55}, 5112-5128 (1997)
doi:10.1201/9781482268737-37
[arXiv:hep-th/9610043 [hep-th]].

\bibitem{Seiberg:1997ad}
N.~Seiberg,
Phys. Rev. Lett. \textbf{79}, 3577-3580 (1997)
doi:10.1103/PhysRevLett.79.3577
[arXiv:hep-th/9710009 [hep-th]].

\bibitem{Aoki:1998bq}
H.~Aoki, S.~Iso, H.~Kawai, Y.~Kitazawa, A.~Tsuchiya and T.~Tada,
Prog. Theor. Phys. Suppl. \textbf{134}, 47-83 (1999)
doi:10.1143/PTPS.134.47
[arXiv:hep-th/9908038 [hep-th]].


\bibitem{Berenstein:2002jq}
D.~E.~Berenstein, J.~M.~Maldacena and H.~S.~Nastase,
JHEP \textbf{04}, 013 (2002)
doi:10.1088/1126-6708/2002/04/013
[arXiv:hep-th/0202021 [hep-th]].


\bibitem{Nicolai:1998ic}
H.~Nicolai and R.~Helling,
[arXiv:hep-th/9809103 [hep-th]].





\bibitem{Dasgupta:2002hx}
K.~Dasgupta, M.~M.~Sheikh-Jabbari and M.~Van Raamsdonk,
JHEP \textbf{05}, 056 (2002)
doi:10.1088/1126-6708/2002/05/056
[arXiv:hep-th/0205185 [hep-th]].


\bibitem{Unsal:2007jx}
M.~Unsal,
Phys. Rev. D \textbf{80}, 065001 (2009)
doi:10.1103/PhysRevD.80.065001
[arXiv:0709.3269 [hep-th]].


\bibitem{Cordova:2018acb}
C.~C{\'o}rdova and T.~T.~Dumitrescu,
SciPost Phys. \textbf{16}, no.5, 139 (2024)
doi:10.21468/SciPostPhys.16.5.139
[arXiv:1806.09592 [hep-th]].


\bibitem{Chen:2020syd}
S.~Chen, K.~Fukushima, H.~Nishimura and Y.~Tanizaki,
Phys. Rev. D \textbf{102}, no.3, 034020 (2020)
doi:10.1103/PhysRevD.102.034020
[arXiv:2006.01487 [hep-th]].

\bibitem{Chen:2024obv}
S.~Chen, E.~Ievlev and M.~Shifman,
Phys. Rev. D \textbf{111}, no.11, 114005 (2025)
doi:10.1103/PhysRevD.111.114005
[arXiv:2411.16845 [hep-th]].



%

\bibitem{Astrakhantsev:2020tdl}
N.~Astrakhantsev, V.~V.~Braguta, E.~M.~Ilgenfritz, A.~Y.~Kotov and A.~A.~Nikolaev,
Phys. Rev. D \textbf{102}, no.7, 074507 (2020)
[arXiv:2007.07640 [hep-lat]].

\bibitem{Begun:2022bxj}
A.~Begun, V.~G.~Bornyakov, V.~A.~Goy, A.~Nakamura and R.~N.~Rogalyov,
Phys. Rev. D \textbf{105}, no.11, 114505 (2022)
[arXiv:2203.04909 [hep-lat]].



\bibitem{Braguta:2023yhd}
V.~V.~Braguta,
Symmetry \textbf{15}, no.7, 1466 (2023)


\bibitem{Iida:2024irv}
K.~Iida, E.~Itou, K.~Murakami and D.~Suenaga,
JHEP \textbf{10}, 022 (2024)
doi:10.1007/JHEP10(2024)022
[arXiv:2405.20566 [hep-lat]].

\bibitem{Das:2025utp}
D.~Das, L.~Ebner, S.~V.~Kadam, I.~Raychowdhury, A.~Sch{\"a}fer and X.~Yao,
[arXiv:2509.18269 [hep-th]].



\bibitem{Wosiek:2002nm}
J.~Wosiek,
Nucl. Phys. B \textbf{644}, 85-112 (2002)
doi:10.1016/S0550-3213(02)00810-6
[arXiv:hep-th/0203116 [hep-th]].

\bibitem{Campostrini:2004bs}
M.~Campostrini and J.~Wosiek,
Nucl. Phys. B \textbf{703}, 454-498 (2004)
doi:10.1016/j.nuclphysb.2004.10.022
[arXiv:hep-th/0407021 [hep-th]].

\bibitem{Anous:2017mwr}
T.~Anous and C.~Cogburn,
Phys. Rev. D \textbf{100}, no.6, 066023 (2019)
doi:10.1103/PhysRevD.100.066023
[arXiv:1701.07511 [hep-th]].


\bibitem{Asplund:2015yda}
C.~T.~Asplund, F.~Denef and E.~Dzienkowski,
JHEP \textbf{01}, 055 (2016)
doi:10.1007/JHEP01(2016)055


\bibitem{Filev:2015hia}
V.~G.~Filev and D.~O'Connor,
JHEP \textbf{05}, 167 (2016)
doi:10.1007/JHEP05(2016)167

\bibitem{Asano:2018nol}
Y.~Asano, V.~G.~Filev, S.~Kov\'a\v{c}ik and D.~O'Connor,
JHEP \textbf{07}, 152 (2018)
doi:10.1007/JHEP07(2018)152

\bibitem{Han:2019wue}
X.~Han and S.~A.~Hartnoll,
Phys. Rev. X \textbf{10}, no.1, 011069 (2020)
doi:10.1103/PhysRevX.10.011069


\bibitem{Garcia-Garcia:2005azc}
A.~M.~Garcia-Garcia and J.~C.~Osborn,
Nucl. Phys. A \textbf{770}, 141-161 (2006)
doi:10.1016/j.nuclphysa.2006.02.011
[arXiv:hep-lat/0512025 [hep-lat]].

\bibitem{Garcia-Garcia:2006vlk}
A.~M.~Garcia-Garcia and J.~C.~Osborn,
Phys. Rev. D \textbf{75}, 034503 (2007)
doi:10.1103/PhysRevD.75.034503
[arXiv:hep-lat/0611019 [hep-lat]].



\bibitem{Hwang:2015}
Myung-Joong Hwang, Ricardo Puebla and Martin B. Plenio, 
Phys. Rev. Lett. \textbf{115}, 180404 (2015)


\bibitem{Larson:2017}
J.~Larson and E.~K.~Irish,
 J. Phys. A: Math. Theor., {\bf 50}(17),174002 (2017).



\bibitem{Beaume:1974ad}
R.~Beaume, J.~Manuceau, A.~Pellet and M.~Sirugue,
Commun. Math. Phys. \textbf{38}, 29-45 (1974)
doi:10.1007/BF01651547


\bibitem{Acerbi_combo}
F.~Acerbi, G.~Morchio, and F.~Strocchi,
 Lett. Math. Phys., {\bf 26}(1), 13-22 (1992); 
J. Math. Phys. \textbf{34}, 899-914 (1993); Lett. Math. Phys.,{\bf 27}(1), 1-11, (1993)


\bibitem{Asorey:1996klg}
M.~Asorey and F.~Falceto,
Phys. Rev. Lett. \textbf{77}, 3074 (1996)
doi:10.1103/PhysRevLett.77.3074
[arXiv:hep-th/9711095 [hep-th]].

\bibitem{Iwai:2010}
T.~Iwai,
J. Phys. A: Math. Theor. \textbf{43} (2010), 415203

\bibitem{Witten:1982df}
E.~Witten,
Nucl. Phys. B \textbf{202}, 253 (1982)
doi:10.1016/0550-3213(82)90071-2


\bibitem{Esteve:1986db}
J.~G.~Esteve,
Phys. Rev. D \textbf{34}, 674-677 (1986)
doi:10.1103/PhysRevD.34.674




\bibitem{Balachandran:2011bv}
A.~P.~Balachandran and A.~R.~de Queiroz,
Phys. Rev. D \textbf{85}, 025017 (2012)
doi:10.1103/PhysRevD.85.025017
[arXiv:1108.3898 [hep-th]].

\bibitem{Neirotti:1998}
J.~P.~Neirotti, P.~Serra, and S.~Kais,
Phys. Rev. Lett., {\bf 80}(24), 5243-5246 (1998)

























\bibitem{ErrastiDiez:2020iyk}
V.~Errasti D\'\i{}ez, M.~Pandey and S.~Vaidya,
Phys. Rev. D \textbf{102}, no.7, 074024 (2020)
doi:10.1103/PhysRevD.102.074024








\bibitem{Braguta:2015owi}
V.~V.~Braguta, E.~M.~Ilgenfritz, A.~Y.~Kotov, B.~Petersson and S.~A.~Skinderev,
Phys. Rev. D \textbf{93}, no.3, 034509 (2016)
doi:10.1103/PhysRevD.93.034509


\bibitem{Braguta:2016aov}
V.~V.~Braguta and A.~Y.~Kotov,
Phys. Rev. D \textbf{93}, no.10, 105025 (2016)
doi:10.1103/PhysRevD.93.105025






\end{thebibliography}
\end{document}